\documentclass[12pt]{article}
\usepackage{feynmp}
\usepackage{amsmath}
\usepackage{amssymb}
\usepackage{latexsym}
\usepackage{graphicx}

\newcommand{\be}{\begin{eqnarray}}
\newcommand{\ee}{\end{eqnarray}}
\newcommand{\half}{\frac{1}{2}}
\renewcommand{\d}{\partial}
\newcommand{\nn}{\nonumber}
\newcommand{\bea}[1]{\left( \begin{array}{#1}}
\newcommand{\ena}{\end{array} \right)}

\newcommand{\CA}{{\cal A}}

\newcommand{\CM}{{\cal M}}

\newcommand{\CO}{{\cal O}}

\newcommand{\CR}{{\cal R}}

\newcommand{\CV}{{\cal V}}

\newcommand\x{\times}
\newcommand\R{\mathbb{R}}
\newcommand\ptl{\partial}
\newcommand\p[1]{\left(#1\right)}
\renewcommand\bar{\overline}
\newcommand\SO{\mathrm{SO}}
\newcommand\SL{\mathrm{SL}}
\renewcommand\>{\rangle}
\newcommand\<{\langle}

\newcommand\G{\Gamma}

\newcommand\w\omega
\newcommand\cL{\mathcal{L}}
\newcommand\z\zeta
\renewcommand\l\lambda
\newcommand\f\phi
\newcommand\De\Delta
\newcommand\s\sigma
\newcommand\oo\infty
\newcommand\g\gamma
\newcommand\vol{\mathrm{vol}}
\newcommand\e\epsilon
\renewcommand\a\alpha
\renewcommand\b\beta
\newcommand\de\delta
\newcommand\pdr[2]{\frac{\ptl #1}{\ptl #2}}

\renewcommand\.{\cdot}
\renewcommand\th\theta

\renewcommand{\tilde}{\widetilde}
\renewcommand{\hat}{\widehat}

\newcommand{\sect}[1]{\section{#1}\setcounter{equation}{0}}

\newcommand\lrpar{\raise .8ex\hbox{$^\leftrightarrow$} \hspace{-9pt}
\partial}
\newcommand\lpar{\raise .8ex\hbox{$^\leftarrow$} \hspace{-9pt}
\partial}
\newcommand\rpar{\raise .8ex\hbox{$^\rightarrow$} \hspace{-9pt}
\partial}

\newcommand{\gsim}{\lower.7ex\hbox{$\;\stackrel{\textstyle>}{\sim}\;$}}
\newcommand{\lsim}{\lower.7ex\hbox{$\;\stackrel{\textstyle<}{\sim}\;$}}
\begin{document}

\baselineskip=18pt

\setcounter{footnote}{0}
\setcounter{figure}{0}
\setcounter{table}{0}

\begin{titlepage}

\begin{center}
\vspace{1cm}

{\Large \bf  Effective Conformal Theory and the Flat-Space Limit of AdS}

\vspace{0.8cm}

{\bf A. Liam Fitzpatrick$^1$, Emanuel Katz$^1$, David Poland$^2$,
David Simmons-Duffin$^2$}

\vspace{.5cm}

{\it $^1$ Department of Physics, Boston University, \\ Boston, MA
02215, USA}

{\it $^2$ Department of Physics, Harvard University, \\ Cambridge, MA
02139, USA}

\end{center}
\vspace{1cm}

\begin{abstract}

We develop the idea of an effective conformal theory describing
the low-lying spectrum of the dilatation operator in a CFT.  Such
an effective theory is useful when the spectrum contains a
hierarchy in the dimension of operators, and a small parameter
whose role is similar to that of $1/N$ in a large $N$ gauge
theory. These criteria insure that there is a regime where the
dilatation operator is modified perturbatively. Global AdS
is the natural framework for perturbations of the dilatation 
operator respecting conformal invariance, much as Minkowski space 
naturally describes Lorentz invariant perturbations of the Hamiltonian.  
Assuming that the lowest-dimension single-trace operator
is a scalar, $\CO$, we consider the anomalous dimensions,
$\gamma(n,l)$, of the double-trace operators of the form $\CO
(\d^2)^n (\d)^l \CO$.  Purely from the CFT we find that perturbative unitarity
places a bound on these dimensions of $|\gamma(n,l)|<4$.
Non-renormalizable AdS interactions lead to violations of
the bound at large values of $n$.  We also consider the case that
these interactions are generated by integrating out a heavy scalar
field in AdS.  We show that the presence of the heavy field 
``unitarizes'' the growth in the anomalous dimensions, and leads to a 
resonance-like behavior in $\gamma(n,l)$ when $n$ is close to the 
dimension of the CFT operator dual to the heavy field.  Finally, we demonstrate
that bulk flat-space S-matrix elements can be extracted from the
large $n$ behavior of the anomalous dimensions.  This leads to a
direct connection between the spectrum of anomalous dimensions in 
$d$-dimensional CFTs and flat-space S-matrix elements in $d+1$ dimensions.
We comment on the emergence of flat-space locality from the CFT perspective.

\end{abstract}

\bigskip
\bigskip

\end{titlepage}

%%%%%%%%%%%%%%%%%%%%%%%%%%%%%%%%%%%%%%%%%%%%%%%%%%%%%%%%%%%%%%%
%\tableofcontents
%\vfill\eject
\sect{Introduction}

One of the central puzzles of the AdS/CFT correspondence~\cite{Maldacena:1997re,Gubser:1998bc,Witten:1998qj}
concerns determining which CFTs have well-behaved AdS descriptions.  A well-behaved description is usually taken to
mean an effective theory containing several AdS fields whose interactions allow a perturbative description over a
range of scales.  Thus, bulk theories typically contain fields whose masses are of order the AdS curvature scale,
while their non-renormalizable interactions are suppressed by a scale much larger than the curvature scale.
In particular, the bulk Planck scale must also be large compared to the AdS curvature scale.
Local bulk scattering of the light fields then satisfies perturbative unitarity until one reaches the scale of
non-renormalizable operators.  Though high-energy scattering appears to violate unitarity, the expectation is that the infinitely many
heavy AdS fields will ultimately ``unitarize'' this scattering, much as QCD resonances lead to sensible scattering of pions.
The low-energy bulk description is therefore valid as long as tree level processes are far from violating the bounds
of perturbative unitarity.

From the AdS effective theory perspective, it appears therefore
that what is essential for the simplicity of description is simply
the existence of a small sector of the theory that is lighter than the
Planck scale and most other states.\footnote{For instance,
supersymmetry does not appear to have a direct role in ensuring
that the bulk effective theory is well behaved, although it might
be important for selecting which low-energy bulk descriptions have
actual UV completions.}  Since the AdS/CFT dictionary relates
dimensions of operators to masses of fields in the bulk, a natural
conjecture, proposed by~\cite{polchinski}, is that any CFT with a
few low dimension operators separated by a hierarchy from the
dimension of other operators will have a well-behaved dual.
However, as any CFT contains an energy-momentum tensor (dual to
the graviton in AdS), there must also be an additional condition
to suppress gravitational interactions in the bulk.  In most known
cases this condition follows from the existence of a large number
of degrees of freedom in the CFT (typically, one takes the
large $N$ limit of an $SU(N)$ gauge theory).  The large $N$ limit
suppresses the connected pieces of higher-point correlation
functions as compared to two-point functions.  $1/N$ thus behaves
as a natural expansion parameter for bulk interactions, and allows
one to distinguish between operators dual to single-particle bulk
states, and those dual to multiple-particle bulk states.  The idea
suggested by~\cite{polchinski} is that having a hierarchy in
dimensions and a parameter such as $N$ in a CFT is sufficient to
construct a sensible AdS effective theory.  This theory describes 
well the correlation functions of low-dimension operators.

A natural question to ask is then what is the CFT interpretation
of the bulk effective field theory.  In particular, there must be an
effective conformal theory (ECT) description which includes only
low-dimension CFT operators as states.  This ECT must be able to distinguish
between renormalizable and non-renormalizable bulk interactions.
It must also obey a condition equivalent to bulk perturbative
unitarity which sets the range of its validity.  Finally,
following standard effective field theory mythology, it would be
satisfying, if in the case that the non-renormalizable terms come
from ``integrating out'' a high-dimension operator with
renormalizable interactions, that perturbative unitarity is
restored on the CFT side.  We will see that the ECT indeed has
these features once we determine the appropriate CFT condition for
perturbative unitarity.

For simplicity, following~\cite{polchinski}, we will consider a
scenario where the lowest-dimension operator is a scalar operator,
$\CO(x)$, with dimension $\Delta$.  We will refer to $\CO(x)$ as a
``single-trace operator" in analogy to large $N$ gauge theories
with adjoint representations, but it is not necessary for the
operator to have this origin.  Other single-trace operators are
taken to have much larger dimensions.  We assume that there is a 
parameter such as $N$ so that at zero-th order in $1/N$ the primary 
operators appearing in the  $\CO \times \CO$ operator product 
expansion (OPE) are the ``double-trace operators'', which have the 
schematic form
\begin{equation}
\CO_{n,l}(x) \equiv  \CO (\lrpar_\nu \lrpar^\nu)^n \lrpar_{\mu_1}...\lrpar_{\mu_l} \CO(x) - \textrm{traces} .
\end{equation}
Here, $\lrpar = \lpar-\rpar$, where the arrows indicate which of
the two operators the derivative acts upon.  At zero-th order in $1/N$ 
the dimension of this operator is given by $2\Delta + 2n + l$.  We 
will be interested in computing the correction to this
dimension, $\gamma(n,l)$, arising from bulk interactions.  For
previous work on computing the anomalous dimensions of double-trace 
operators in the context of AdS/CFT, see e.g.~\cite{Liu:1998th,D'Hoker:1998mz,D'Hoker:1999pj,D'Hoker:1999jp,Hoffmann:2000mx,Arutyunov:2000ku,Hoffmann:2000tb,Arutyunov:2002rs,Cornalba:2006xm,Cornalba:2007zb,polchinski,Heemskerk:2010ty}.

In order to develop an ECT, we need to specify a notion of energy
in the CFT that will map nicely to energies in the bulk theory. As
the ECT is supposed to describe low-dimension operators, a natural
notion of energy is the dimension itself.  The Hamiltonian for
which we are developing the ECT is the dilatation operator, and
the ECT is then intended to capture its low-lying spectrum.  In
that sense, for fixed spin, one can think of energy, $E$, as $E
\sim 2 n$. It will be important to keep in mind that this notion of energy
corresponds to the dimensions of CFT operators and is distinct from Poincar\'e 
energy.   From the CFT perspective, the task is to start from a dilatation
operator, $D^{(0)}$, whose spectrum contains a hierarchy, and
perturb it by adding a small correction, $V$, suppressed by $N$.
The new dilatation operator, $D=D^{(0)} + V$, is taken to act on 
the low-dimension sector of $D^{(0)}$.  In our simplified scenario, this
includes multi-trace operators containing only $\CO$ and derivatives.
Calculating $\gamma(n,l)$  thus amounts to diagonalizing $D^{(0)} + V$ in
perturbation theory.  Purely from the CFT, we will show that
perturbative unitarity places a bound on the anomalous dimensions
of $|\gamma(n,l)| < 4$.  We will then turn to calculating the anomalous
dimensions for particular choices of $V$, corresponding to local bulk
interactions in AdS.  For such calculations we find it most natural to work 
in global AdS, since the energy conjugate to
global time is associated with the dilatation operator.  Indeed, we will show that local bulk interactions in global AdS
automatically lead to a $V$ which is consistent with conformal symmetry.
We will then demonstrate that using old-fashioned perturbation theory in
global AdS gives a very efficient method of computing the
anomalous dimensions $\gamma(n,l)$.  This is because 
these anomalous dimensions are just the correction to
the energy in global coordinates\footnote{The Hamiltonian of AdS in global
coordinates is more useful for our purposes than the Hamiltonian in the 
Poincar\'e patch.
This is because translations in global AdS time correspond to dilatations in 
the CFT, whereas time in the Poincar\'e patch corresponds to Poincar\'e time 
in the CFT. } of two-particle AdS states due to bulk interactions.   
Previously, obtaining $\gamma(n,l)$, involved 
extracting the anomalous dimensions from four-point correlation functions
using sophisticated techniques limited to even CFT dimensions.
Our method is simpler and applies for any dimension.

As expected from AdS, the above unitarity bound will be violated
by terms in $V$ coming from non-renormalizable bulk interactions.
Indeed, as would follow from the above identification of $n$ with
energy, we find that a local bulk term suppressed by $\Lambda^p$,
will lead to a growth in $\gamma(n,l) \sim n^p$.\footnote{This
growth was found earlier by~\cite{polchinski} using other
methods.} Thus, the value of $n$ at which the bound is violated
sets a natural boundary for the validity of the ECT.  The
existence of a useful ECT description is then the statement that
perturbative unitarity is not violated over a wide range of $n$'s.
This is related to locality of interactions which include only the
field dual to operator $\CO$ in the bulk theory.

To make connection with the conjecture of~\cite{polchinski}, and
to verify standard effective theory lore, we also consider the
generation of non-renormalizable bulk interactions via the
exchange of a heavy scalar, dual to a CFT operator $\CO_{\rm
Heavy}$ (where $\Delta_{\rm Heavy} \gg \Delta$).  At $n \ll
\Delta_{\rm Heavy}$ we reproduce the exact contributions to
$\gamma(n,l)$ one would expect from the leading non-renormalizable
interactions generated by integrating out the heavy state,
suppressed by the appropriate powers of $\Delta_{\rm Heavy}$. This
result is suggestive that a hierarchy in the dimension of
operators leads to a large range for the ECT.  This example also
shows explicitly how putting a large-dimension operator back into
the ECT ``unitarizes'' $\gamma(n,l)$.  In fact, just as one would expect
from effective field theory, we will see that the growth in $\gamma(n,l)$ 
turns into a resonance at $n \sim \Delta_{\rm Heavy}/2$, before decreasing 
at large $n$.

At energies much larger than the inverse AdS radius it is expected
that one can make contact with flat-space scattering amplitudes.
An important goal that has been pursued using a variety of
methods~\cite{Polchinski:1999ry,Susskind:1998vk,Balasubramanian:1999ri,Giddings:1999qu,giddingslocality,Jevicki:2005ms,Gary:2009ae,Gary:2009mi,Okuda:2010ym}
is to understand how these amplitudes arise from CFT data.  Here
we will show that it is in fact possible to extract the flat-space
S-matrix elements of the bulk theory from the large $n$ behavior
of $\gamma(n,l)$.  
Stated simply, we will argue that at leading order for bulk $\phi$-particle scattering,
\begin{equation}
\CM(s,t,u)^{d+1}_{\textrm{flat space}} \sim \frac{E_n}{(E_n^2-4\Delta^2)^{\frac{d-2}{2}}} \sum_l [\gamma(n,l)]_{n \gg l} ~r_l P^{(d)}_l\left(\cos \theta \right),
\end{equation}
where $r_l P^{(d)}_l\left(\cos \theta\right)$ are the appropriate
polynomials in $d$-dimensions, the total flat-space energy,
$E_n$, is given in units of the AdS radius 
by $E_n=2\Delta+2n$, and $[\gamma(n,l)]_{n \gg l}$ indicates that one needs to 
take the large $n$ limit of $\gamma(n,l)$, keeping $l$ fixed.  In other words, the
$\gamma(n,l)$'s form the partial wave expansion of the higher
dimensional flat-space S-matrix.\footnote{This sharpens the relation
between $\CM$ and $\gamma$ found previously for local bulk operators and neglecting
mass terms
\cite{polchinski}. }  By ``flat-space S-matrix'', one means simply
the scattering amplitudes one obtains from the Lagrangian of the
bulk theory, but applied in Minkowski space.  
It is interesting that there seems
to be such a direct connection between CFT quantities and
flat-space matrix elements.  Note that this connection is only
possible if the ECT including $\CO$ and $\CO_{\rm Heavy}$ obeys
perturbative unitarity for $n$ sufficiently large.  Therefore, a
hierarchy in dimensions and a parameter such as $N$ are essential
for flat space to emerge.

This paper is organized as follows.  In section 2 we will introduce the general formalism 
concerning perturbations of the dilatation operator and discuss the constraints arising 
from perturbative unitarity.  We will then review the construction of scalar wavefunctions 
in global AdS, and discuss why local bulk interactions lead to a sensible perturbation of 
the dilatation operator.  In section 3 we will derive the general form of the wavefunctions 
corresponding to primary operators in the CFT, and use this to calculate the anomalous dimensions 
of primary double-trace operators arising from various bulk quartic interactions.  In section 4, 
we will consider integrating out a heavy scalar field in AdS, and we will compare the resulting 
anomalous dimensions to the leading-order contributions from the low-energy effective field theory.  
In section 5, we will explore the flat-space limit of AdS, and show how flat-space S-matrix elements 
can be determined from the large $n$ behavior of the anomalous dimensions.  We conclude in section 6.

\sect{Formalism}

\subsection{Algebra Constraints}
\label{sec:algebra}

In quantum field theory, free fields provide a fundamental starting point
for perturbation theory because they have a solvable Hamiltonian
and simple dynamics corresponding to multi-particle states.
In conformal field theory, the dual role is played by ``mean fields'',
which have a Gaussian partition function and a simple spectrum of operator
dimensions corresponding to multi-trace operators.  For CFTs arising
from a gauge group with a large rank $N$, corrections to three- and higher $n$-point
correlation functions of canonically normalized primary operators
are expected in general to be suppressed by powers of $N$, so that
the mean field theory correlation functions are a good approximation.
In this case, the dilatation operator $D$ of the CFT may be split
into a mean-field piece $D^{(0)}$ that survives as $N$ is taken to infinity,
and a perturbation $V \equiv D-D^{(0)}$ that is suppressed by some power of $N$.
In radial quantization, where one studies radial evolution rather than
time evolution of the CFT, $D$ plays the role of a Hamiltonian, and so
$V$ plays the role of an interaction.
However, this procedure is not limited to CFTs arising from large-rank
gauge groups; we may perform perturbation theory in this way
any time the CFT reduces to a mean field theory when some small parameter or
parameters vanish.  Thus, we shall follow \cite{polchinski} and use
$N$ in this more general sense, as the formal parameter suppressing
$V$.  Of course, we are not interested in general perturbations
around mean field theory, but rather only in those where the perturbed
theory is also conformal.
A great strength of AdS/CFT is that local AdS-Lorentz invariant
interactions generate perturbations in the CFT of exactly this
form.

We will write the conformal algebra as
\be\label{eq:confalgebra}
\left[ M_{\mu\nu} , P_\rho \right] = i (\eta_{\mu\rho} P_\nu - \eta_{\nu\rho} P_\mu)
, && \left[ M_{\mu\nu} , K_\rho \right] = i (\eta_{\mu\rho} K_\nu  - \eta_{\nu\rho}
K_\mu) , \nn\\
\left[ M_{\mu\nu} , D \right] = 0 , && \left[ P_\mu, K_\nu \right] =
- 2 (\eta_{\mu\nu} D + i M_{\mu\nu} ) , \nn\\
\left[ D, P_\mu \right] =  P_\mu , && \left[ D, K_\mu \right] = - K_\mu .
\ee
Note that we have chosen our convention for $D$ so that it is Hermitian, which
differs from the most common convention by a factor of $i$.  The requirement that
this algebra is held fixed is then a non-trivial constraint on the form of possible
perturbations to the generators.

For simplicity, we will start by specializing to the case of 2d CFTs, where the
algebra can be divided into left and right pieces using the decomposition $\SO(2,2)=\SL(2,\R)_L\x \SL(2,\R)_R$.  In particular, the generators
$M_{\mu\nu}, P_\mu, K_\mu,D$ of the algebra are all linear combinations
of operators that act non-trivially on left-moving states only and
right-moving states only
\be
&& K = \frac{K_1+i K_2}{2}, \ \ \ P = \frac{P_1 - i P_2 }{2}, \ \ \ L_0 = \frac{D-M_{12}}{2},
  \ \ \ \textrm{(left-moving)}, \nn\\
&& \bar{K} = \frac{K_1-i K_2}{2} , \ \ \ \bar{P} = \frac{P_1 + i P_2}{2}, \ \ \
\bar{L}_0 = \frac{D+M_{12}}{2}, \ \ \ \textrm{(right-moving)}.
\ee
The left-moving generators then satisfy the algebra
\be
&& [L_0, P] = P , \ \ [L_0, K] = - K , \ \ [P,K] = - 2 L_0 ,
\ee
and the right-moving generators satisfy the same algebra, with $K,P,L_0 \rightarrow
\bar{K}, \bar{P}, \bar{L}_0$.

Focusing on the left-moving algebra, we can now split the generators into mean field theory
generators and $O(1/N^2)$ corrections.  In general, the perturbations will
be constructed so that $M_{\mu\nu}$ is unmodified, so that both $L_0$
and $\bar{L}_0$ get corrected by $\half V$:
\be
L_0 &=& L_0^{(0)} + \frac12 V, \nn\\
P   &=& P^{(0)} + P^{(1)}, \nn\\
K   &=& K^{(0)} + K^{(1)}.
\ee
The constraint that the theory is still conformal 
then implies the following relations at $O(1/N^2)$
among the perturbations to the generators:
\be
\left[ \frac12 V, K^{(0)} \right] + \left[  L_0^{(0)}, K^{(1)} \right] 
 &=& - K^{(1)} ,\nn\\
\left[ \frac12 V, P^{(0)} \right] + \left[ L_0^{(0)}, P^{(1)} \right]
 &=& P^{(1)} ,\nn\\
\left[ P^{(1)} , K^{(0)} \right] + \left[ P^{(0)}, K^{(1)} \right]
 &=& - V .
\label{eq:expandedcommute}
\ee
These relations turn out to be extremely useful.  To derive their implications
for the matrix elements of the perturbed generators, let us
choose our basis to be the eigenstates of $L_0^{(0)}$.
As usual, the left-moving states are classified as primary states, which are
annihilated by $K$, or descendant states, which are obtained from
the primary states by acting repeatedly with $P$.  In this subsection, we
will denote a primary state with $L_0^{(0)}$ eigenvalue $\alpha$ as
$|\alpha, 0\>$, and its normalized $m$-th descendant as
$|\alpha, m\>$.
It is then straightforward using the algebra to work out the
action of the zero-th order generators on any state.  In particular,
\be
L_0^{(0)} | \alpha, m \> &=& (\alpha+m)| \alpha, m\>, \nn\\
P^{(0)} |\alpha, m \> &=& \sqrt{(m+1)(2\alpha+m)}|\alpha, m+1\>
  \equiv c^\alpha_m | \alpha, m+1 \>, \nn\\
K^{(0)} | \alpha, m \> &=& \sqrt{m(2\alpha+m-1)}
|\alpha, m-1\> = c^\alpha_{m-1} | \alpha, m-1 \> .
\ee
By taking matrix elements of Eqs.~(\ref{eq:expandedcommute})
between zero-th order states $\< \alpha,m |$ and $| \beta,m' \>$,
we obtain three separate equations.  The first can be written as
\be
2 K^{(1)}_{\alpha,m; \beta,m'} &=&
  \frac{c^\alpha_m V_{\alpha,m+1; \beta,m'} - c^\beta_{m'-1} V_{\alpha,m;
\beta,m'-1}}{1+\alpha+m-\beta-m'},
\label{eq:kperturbation}
\ee
where $\CO_{\alpha,m; \beta,m'}$ denotes $\< \alpha, m | \CO | \beta, m' \>$.
The second condition in Eq.~(\ref{eq:expandedcommute}) becomes
\be
2 P^{(1)}_{\alpha,m; \beta,m'} &=& \frac{c^\alpha_{m-1} V_{\alpha,m-1; \beta,m'}
 - c^\beta_{m'} V_{\alpha,m; \beta, m'+1}}{-1+\alpha+m-\beta-m'},
\ee
which follows from the first one using $P = K^\dagger, V=V^\dagger$.
The third condition of Eq.~(\ref{eq:expandedcommute}) also follows from the first two.
Thus, all of the perturbed generators can be determined from the matrix elements
of the dilatation operator.  One of our major goals will be to calculate and
study the behavior of these matrix elements.

The above relations will be extremely important when we use
time-independent perturbation theory to construct the dilatation eigenstates
at first order.  Na\"{i}vely, a straightforward construction is impossible in
practice because of the enormous zero-th order degeneracy between multi-trace
states.
Thus, one would expect to have to diagonalize $V$ within the space of
degenerate states, which would be intractable for the vast majority
of states of interest.

Fortunately, this is not the case, a fact that follows from the above relations
under the assumption that
$K,P$ have finite matrix elements between zero-th order
dilatation eigenstates.\footnote{This assumption is satisfied by
perturbations generated by local interactions in AdS, except at
particular fractional values of $\alpha$.} Specifically, taking $m'=0$
in Eq.~(\ref{eq:kperturbation}) we see that matrix elements of $V$ between
a primary state $|\beta,0 \>$ and a descendant $|\alpha, m+1\>$ with the
same dimension must vanish!  The reason is that $V_{\alpha,m; \beta,-1}$
must vanish since $|\beta, 0\>$ is primary, and the denominator
$1+\alpha+m-\beta$ also vanishes under the assumption that that the states
have the same dimension.  Thus there is no possible cancellation between
the two terms in the numerator, and since $K^{(1)}$ is assumed to be finite,
we necessarily have $V_{\alpha, m+1; \beta,0} = 0$.  This is very useful,
since it means that we do not have to do degenerate perturbation theory in order
to construct the first-order primary states.

It will be helpful to discuss the space of states further, and to establish
some more notation.  We will be focusing on the simplest
possible CFTs, where the only single-trace primary operator is
a scalar operator $\CO$ with dimension $\Delta$.  Following~\cite{polchinski},
we will be ignoring the role of the energy-momentum tensor $T_{\mu\nu}$ in
the majority of our analysis, which formally corresponds to taking the limit
of very large central charge $c \gg N$.
In a sense, therefore, we will be studying toy models, though we believe
our results are rather general and would apply to theories with a
$T_{\mu\nu}$ as well.
Out of $\CO$, one can make many double-trace primary operators.  In
mean field theory, one knows their form explicitly.  Adopting the notation
of \cite{polchinski}, they are schematically
\be
\CO_{n,l} &=& \CO \lrpar_{\mu_1} \dots \lrpar_{\mu_l}
(\lrpar_\nu \lrpar^\nu)^n \CO - \textrm{traces},
\ee
and they have dimension $E_{n,l} = 2 \Delta + 2n + l$ and spin $l$.
Inserting one of these operators at the origin creates a double-trace
primary state $\CO_{n,l}(0) | 0 \> =  |n,l\>_2$, which we
will label by their $n$ and $l$ values.

When we perturb the mean-field theory dilatation operator by an
interaction $V$, the eigenstates of the perturbed dilatation operator
acquire the anomalous dimensions
\be
\Delta_{n,l} &=& E_{n,l} + \gamma(n,l).
\ee
It is then relatively straightforward to calculate $\gamma(n,l)$ using old-fashioned perturbation theory
\be
\gamma(n,l) &=& {}_2\<n,l| V |n,l\>_2 + \sum_{\alpha} \frac{|\<\alpha| V |n,l \>_2|^2}{E_{n,l}-E_{\alpha}} + \dots,
\ee
where $E_\alpha$ is the leading order dimension of $|\alpha\>$.
In this paper we will give a number of concrete examples which demonstrate how to calculate $\gamma(n,l)$ using the above method.

Of course, not every choice of $V$ will lead to a well-behaved perturbative expansion for all $n$ and $l$.  This is quite similar to the statement that not every interaction in flat space leads to a perturbatively calculable S-matrix for all choices of external energy.  In particular, non-renormalizable interactions
lead to a violation of perturbative unitarity when scattering at sufficiently high
energies.  In the next subsection we will show that large $N$ CFTs have a similar
constraint from perturbative unitarity, which can be stated quite simply in terms
of the large $n$ behavior of $\gamma(n,l)$.  We will later show that this constraint is
satisfied if $V$ arises from renormalizable local bulk interactions in AdS, and is violated
if $V$ arises from non-renormalizable bulk interactions.

\subsection{Unitarity Limit}
\label{sec:unitarity}

The requirement that scattering amplitudes in flat-space field theory
be unitary means that contributions from higher-dimensional operators
cannot continue to grow indefinitely, and eventually the validity
of the effective theory breaks down.  One expects that before this happens,
heavy fields will appear to unitarize the theory.  The systematic
description of such constraints is through the optical theorem, and more
generally through the cutting rules, which will appear to be violated at
tree-level if one considers sufficiently high energy scattering.
It was demonstrated in \cite{polchinski} that all $O(1/N^2)$
CFT perturbations that satisfy crossing symmetry can be generated
by local operators in AdS.
Most of these AdS operators will be non-renormalizable, and we would like to
derive something like an optical theorem which is violated
by conformal theories with perturbations generated
by higher-dimensional operators in AdS.  Na\"{i}vely, there can be no such limit.
At tree-level, a generic AdS action essentially defines a CFT at $O(1/N^2)$,
and the correlation functions are perfectly well-behaved.  Indeed, since
there is no scale in the CFT, there would appear to be ipso facto
no scale where the theory could break down.  However, the point is that the
CFT secretly does have something that plays the role of a scale:
the $n$ in the double-trace primary operators $\CO_{n,l}$.

By considering a scattering thought experiment in AdS
and relating it to CFT correlation functions,~\cite{polchinski}
found that the anomalous dimension $\gamma(n,l)$ of $\CO_{n,l}$
generated by a non-renormalizable interaction in AdS$_{d+1}$ of scaling dimension
$p$ must grow like $n^{p-(d+1)}$.  As a result, regardless of how small
$1/N$ is, for $p > d+2 $ there will be some $n$ above which the $O(1/N^2)$
corrections to the dimension of a double-trace primary operator is larger
than the leading term $2\Delta + 2n +l$.\footnote{We thank Jo\~{a}o Penedones for pointing this out to us,
and for noting that the dimensions of double-trace operators 
will become negative if the sign of the AdS interaction is chosen incorrectly.}
  Our goal in this section will be to
find a sharp limit where this growth leads to problems, and in the process
tighten the constraint to apply to non-renormalizable operators with
$p > d+1$.

We can try to set up something like an optical theorem in terms of
CFT quantities. The dilatation eigenstates $|A\>$
of the perturbed theory will be related to those of the unperturbed
theory through a transition matrix $T$
\be
| A \> &=& \left( \delta_{AB} + T_{AB} \right) | B \>^{(0)} .
\label{eq:Tdef}
\ee
The optical theorem in quantum field theory follows just from completeness
of the ``in'' and ``out'' states, and the fact that the S-matrix is just
a change of basis. The most similar condition we can build out of the
CFT quantities at hand is the completeness of the perturbed and unperturbed
eigenstates
\be
\delta_{AB} &=& \sum_C \< A | C\>^{(0)} {}^{(0)}\< C| B \> .
\ee
Here, it is important to note that we will be interested in applying
this completeness relation to the low-lying states of the dilatation operator.
Indeed, changing $N$ in the full CFT will in general modify the Hilbert space
and therefore the eigenstates of $D$ and $D^{(0)}$ are not strictly describing the
same space.\footnote{We thank Joe Polchinski for bringing up this point.}  However, 
at large $N$ it will be the large-dimension operators (with dimensions of $O(N)$) that will 
be sensitive to such changes in the Hilbert space, not the low-dimension ones.
This is very similar to the situation in large $N$ QCD where one is similarly changing
the Hilbert space by varying $N$.  At large $N$, however, the subspace of low-mass meson
states (of mass, $m \ll N \Lambda_{QCD}$) is not changing significantly. 
In fact, perturbative unitarity of the S-matrix is precisely the criterion one uses to determine the 
range of energy and mass over which a change in $N$ is not modifying the space of states.

Let us then find the implication of the above completeness relation.  
If we insert (\ref{eq:Tdef}) and take $A=B$, we find
\be
- (T+T^*)_{AA}  &=& \sum_C |T_{AC}|^2 .
\ee
It is clear from this relation that $\CR e(T)_{AA} < 0$, which one
should keep in mind in the following manipulations.  Using
that $\sum_C |T_{AC}|^2 > |\CR e(T)_{AA}|^2$, one obtains the constraint
on $|\CR e(T)_{AA}|$ that
\be
| \CR e(T_{AA}) | &<& 2 .
\ee
This limit must be satisfied, and we will refer to it as the
unitarity limit since it followed from the fact that
$\< A | C \>^{(0)}$ is a unitary matrix. Consider now the condition
that it be satisfied in perturbation theory.  The first
contribution to $\CR e (T_{AA})$ occurs at $O (V^2)$ from
the renormalization $|A\> \rightarrow Z_A^{-\half} |A \>$, where 
$Z_A = 1 + \sum_{B\ne A} |V_{AB}|^2/(E_A-E_B)^2 + O(V^3)$ and $E_A$ 
denotes the zero-th order dimension of $|A\>$.
Thus, at $O(V^2)$, we have
\be
&& 2> | \CR e (T_{AA})| =  \half \sum_{B\ne A} \frac{ |V_{AB}|^2}
{(E_A - E_B)^2}.
\ee
Let us now take
$| A \>, |B\>$ to be neighboring double-trace primary states $|n, l \>_2$
and $|n+1, l \>_2$, respectively.  The difference in their mean field
dimensions is exactly 2, so the above relation implies
$|V_{n,l;n+1,l} | < 4$, since every other term on the right hand side is positive.
At large $n$, there is not much difference between $V_{n,l;n+1,l}$ and $V_{n,l;n,l}$.
Both can be calculated from the overlap of wavefunctions in AdS, and the
difference between wavefunctions for $|n,l \>_2$ and $|n+1,l\>_2$ is
$O(1/n)$ at large $n$.  This will become
especially obvious when we consider example calculations of matrix elements
of $V$. But, $V_{n,l;n,l}$ is just the leading order
anomalous dimension $\gamma(n,l)$ of
the state $|n,l \>_2$. Thus, we can state a very simple necessary condition
in order to maintain perturbative unitarity in the CFT $1/N^2$ expansion
\be
\phantom{.         .} |\gamma(n,l)| &<& 4 \ \ \ (n \gg 1)
\ee
What this says is that perturbation theory fails when the anomalous
dimensions $\gamma(n,l)$ become much greater than 1.\footnote{We note that the above bound is not as general as those derived in~\cite{Rattazzi:2008pe,Rychkov:2009ij,Caracciolo:2009bx}, which are valid also when both $n$ and $N$ are small, and are thus non-perturbative statements.}
In fact, tracing back the steps that lead to this break-down, we see that
the states $|n,l \>_2$ have negative norm at $O(1/N^2)$ when the above
condition is not satisfied. When this happens, the description of the
CFT must be modified to maintain unitarity, and if this is to occur
before the $n$ where perturbation theory fails then one must have
new large-dimension single trace operators that contribute to
$\gamma(n,l)$ and unitarize the transition matrix.  Even if new
single-trace operators do not appear before $|\gamma(n,l)|>4$, the theory
becomes ``strongly coupled'' at that point, in the sense that $V$ is large, and
the standard lore is that the modified description of the
theory at large $n$ will contain additional heavy states.

Consequently, the implications of large $n$ growth are fairly striking.
Na\"{i}vely, effective field theories in AdS are dual to a very limited class
of CFTs.  In order for the AdS EFT to be calculable, all non-renormalizable
operators must be suppressed at least by appropriate powers of some
scale $\Lambda$, the cut-off of the theory. For example, consider all possible
four-point contact interactions of a scalar field $\phi(x)$, dual to a CFT
operator $\CO$.
Such four-$\phi$ interactions are in one-to-one correspondence
with all different possible crossing-symmetric contributions
to the $\CO$ four-point
function~\cite{polchinski}.  Thus, we appear to require an infinite
number of conditions on the CFT four-point function, one for each
non-renormalizable operator in AdS.  What the above discussion says
is that all of these apparently independent conditions are simply the
condition of a hierarchy in the dimensions of operators
in the  CFT, with no new single-trace primary operators appearing below some
dimension $\Delta_{\rm Heavy}$. Furthermore, the suppression of the
perturbations dual to non-renormalizable AdS interactions is given by
appropriate powers of $\Delta_{\rm Heavy}$.
This is exactly dual to the condition in AdS that there
is a hierarchy in scales between the mass of $\phi$ (and whatever other
fields appear in our effective theory) and the new physics that appears
around the cut-off $\Lambda$. In the following sections, we will explore
this relation further, and in particular the description within
the CFT of the transition at low $n$ below $\Delta_{\rm Heavy}$ to
large $n$, where the heavy conformal sector is ``integrated in'' to
restore unitarity.

\subsection{Review of AdS Global Coordinate Wavefunctions}

Next we will turn to the concrete construction of effective field
theories in AdS. The connection between fields in AdS and
operators with definite scaling dimension in the CFT is
significantly more transparent in global coordinates than in
Poincar\'e coordinates.  For completeness and to establish
notation, we will now review this connection in detail
\cite{bf,Witten:1998qj}, as well as the construction of the
canonical field operators in AdS global coordinates.

To begin, we work in global coordinates in AdS$_{d+1}$, with the metric
\be
ds^2 &=& \frac{1}{\cos^2 \rho} \left( -dt^2 + d\rho^2
  + \sin^2 \rho\, d \Omega^2\right) .
\ee
We will work in units of the AdS radius $R_{\rm AdS} \rightarrow 1$.  The center of AdS lies at
$\rho=0$, and the boundary at $\rho=\pi/2$.  The boundary manifold
is $\R \times S^{d-1}$, where translations in global coordinate time
generate dilatations in the CFT.

We will now consider a bulk scalar field $\phi(x)$, dual to a single-trace
scalar operator $\CO(0)$ and its descendants in the boundary CFT.  The free field wavefunctions in AdS$_{d+1}$
satisfy $(\nabla^2- m^2)\phi =0$, which has the
solutions (keeping only the modes which are well-behaved at $\rho=0, \pi/2$)
\be
\phi_{nlJ}(x) &=& \frac{1}{N_{\De,n,l} } e^{i E_{n,l} t} Y_{lJ}(\Omega)
  \sin^l\rho \cos^\Delta\rho F(-n, \Delta + l +n, l+ \frac{d}{2},
\sin^2 \rho) \nn\\
E_{n,l} &\equiv & \Delta + 2n +l, \ \ \ \ m^2 = \Delta (\Delta - d),
\label{eq:singleparticlewavefunctions}
\ee
where
$F=\phantom{}_2F_1$ is the Gauss hypergeometric function, $Y_{lJ}(\Omega)$ are normalized
eigenstates of the Laplacian on $S^{d-1}$ with eigenvalue $-l(l+d-2)$,
and $J$ denotes all the angular quantum numbers other than $l$.
In many formulae, dependence on the $J$ index will be clear from context 
and we will often suppress it.  
The canonical field operators are then constructed in terms of the
wavefunctions and creation/annihilation operators
\be
\phi(x) &=& \sum_{n,l,J} \phi_{nlJ}(x) a_{nlJ} + \phi^*_{nlJ}(x) a^\dagger_{nlJ} \ .
\ee
We will denote the one-particle states created by $a^\dagger_{nlJ}$ 
as $|\phi ; n, l, J \>$, where
indices after the semi-colon indicate descendants.  They are in
one-to-one correspondence with the states created at the origin by the
single-trace operator $\CO(0)$ and its descendants, since both are
simply the eigenstates of the dilatation and rotation operators with
energy $\Delta + 2n +l$.  This is what makes AdS global coordinates a
natural place to work when studying anomalous dimensions of operators.

Using the norm $(\phi_1, \phi_2) \equiv  \int d^d x \sqrt{-g} g^{00} \phi_1(x)^*
\lrpar_0 \phi_2(x)$, the wavefunctions are properly normalized when
\be
N_{\De,n,l} &=& (-1)^n \sqrt{  \frac{n! \Gamma^2(l+\frac{d}{2}) \Gamma(\Delta + n - \frac{d-2}{2}) }
{ \Gamma(n+l+\frac{d}{2}) \Gamma(\Delta + n + l)}} ,
\label{eq:singleparticlenorm}
\ee
where we have chosen the $n$-dependent phase for later convenience.

In addition to the one-particle wavefunctions, we will be interested in more general wavefunctions
(e.g., two-particle wavefunctions) in AdS that are dual to primary states in the CFT.  In order to study this
we will need to understand the action of the conformal generators on functions of AdS global coordinates.
This is most easily determined by going to the embedding space of AdS$_{d+1}$, which we will write as
\be
ds^2 &=& - dX_0^2 - dX_{d+1}^2 + \sum_{\mu=1}^d dX_{\mu}^2, \ \ \ \ \
- 1 = X_M X^M .
\ee
The embedding space coordinates are then related to global coordinates through the identifications
\be
X_0 &=& \frac{\cos t}{\cos \rho}, \ \ \ \ \ X_{d+1} = \frac{\sin t}{\cos \rho}, \ \ \ \ \ X_{\mu} = \tan \rho\,\Omega_{\mu} .
\ee
The generators of the $SO(d,2)$ algebra are simply represented in the embedding space as
$J_{M N} = -i (X_M \d_N - X_N \d_M)$.  In particular, the conformal algebra Eq.~(\ref{eq:confalgebra})
is correctly reproduced if we identify
\be
\label{eq:conformalgenerators}
&& P_\mu = J_{\mu, d+1} - i J_{\mu, 0} \hspace{1cm}
K_\mu =  J_{\mu, d+1} + i J_{\mu,0 }
\hspace{1cm} D = - J_{0,d+1} \hspace{1cm} M_{\mu\nu} =  J_{\mu\nu} .
\ee
It is then straightforward to work out their corresponding action in
terms of global coordinates.  For example, in general $D=-i \d_t$, and
 in AdS$_3$ the left- and right-moving generators act as
\be
K_{\pm}
  &=&  i e^{-it \pm i \varphi} \left( \sin \rho \d_t + i \cos \rho \d_\rho
  \mp \frac{1}{\sin \rho} \d_\varphi \right) \nn\\
P_{\pm}
  &=&  i e^{it \pm i \varphi} \left( \sin \rho \d_t - i \cos \rho \d_\rho
  \pm \frac{1}{\sin \rho} \d_\varphi \right)
  \label{2dgen}
\ee
where $K_\pm = K_1 \pm i K_2, P_\pm = P_1 \pm  i P_2$.

We are now in position to construct the wavefunctions in AdS that are dual
to the double-trace primary operators $\CO_{n,l}(0)$.  We will do this in detail in section~\ref{sec:dilatationmatrix}.
Afterwords we will consider adding local bulk interactions $\CV(x)$, treating $V = \int d^d x \CV(x)$ as a perturbation
to the dilatation operator of the CFT.  We will then use old-fashioned perturbation theory in order to calculate the
corrections to the anomalous dimensions $\gamma(n,l)$ arising from $V$.  However, first we would like to consider more carefully why
the integral of a local bulk interaction in AdS leads to a sensible perturbation of the dilatation operator in the dual CFT.

\subsection{Locality and Microcausality in AdS}

In the case of a Lorentz invariant theory in flat space, it is well known that if the interaction part of the Hamiltonian, $V$, can
be written in terms of local interaction density $\CV(x)$ integrated over space, then Lorentz invariance requires that
$[\CV(x),\CV(y)]=0$ for $(x-y)^2 < 0$.  Thus, in order to build Lorentz-invariant interactions for a particular particle, the standard procedure is to take the creation and annihilation operators for that particle and assemble them into a field $\phi(x)$.
 $\phi(x)$ transforms simply under Lorentz transformations, and in addition obeys $[\phi(x),\phi(y)]=0$ for $(x-y)^2 < 0$.
We then build $\CV(x)$ as a scalar operator made up of $\phi(x)$ and its derivatives, $\CV(x) = \CV(\phi(x),\d_\mu\phi(x),\d_\mu\d_\nu\phi(x),...)$.  Such a $\CV(x)$ automatically obeys microcausality and leads to a Lorentz-invariant theory.

In many ways, the procedure in AdS is similar to the Lorentz-invariant case.
We are interested in constructing the interaction part of the dilatation operator, $V$, in a way which gives a conformally
invariant theory.  In the previous section we reviewed how to assemble the creation and annihilation operators
associated with a primary operator in the CFT and its descendants
into an AdS field, $\phi(x,t)$. (Note that here $x$ denotes all coordinates other than the global time $t$.)
Under the AdS isometries $\phi(x,t)$ transforms in a simple way, and it also obeys $[\phi(x,t),\phi(y,t)]=0$ for $x \neq y$
by construction.  If we now build $\CV(x,t)$ as an AdS scalar made out of $\phi(x)$ and its derivatives, it will
also obey $[\CV(x,t),\CV(y,t)]=0$.  We will now check that the AdS microcausality condition on $\CV(x,t)$ is sufficient
to insure that $D=D^{(0)} + V $ is a sensible dilatation operator.   Along the way, we will see explicitly that the
operator $K^{(1)}$ has non-singular matrix elements as discussed in section~\ref{sec:algebra}.

We will make our argument for the case of AdS$_3$ for simplicity, although it naturally generalizes to higher dimensions.  Let
$V = \int d^2x\sqrt{-g} ~\CV(x)$, where $\CV(x)$ is a local interaction density.  Then the leading order special
conformal transformation, $K^{(0)}$, acts on the scalar $\CV(x)$ simply through the corresponding isometry (\ref{2dgen}) of AdS
\be
[K^{(0)}_\pm, \frac{V}{2}] &=&  -\frac{i}{2} \int d^2x\sqrt{-g} ~ e^{-it \pm i \varphi} \left( \sin \rho \d_t + i \cos \rho \d_\rho \mp \frac{1}{\sin \rho} \d_\varphi \right)\CV(x,t).
\ee
Here, $\CV(x,t)$ is evolved using $D^{(0)}$, and so $\d_t \CV(x,t) = -i [D^{(0)}, \CV(x,t)]$.  Consequently, upon integrating the
last two terms in the above expression by parts, one obtains
\be
[K^{(0)}_\pm, \frac{V}{2}] &=&  -\frac{1}{2} \int d^2x\sqrt{-g} \sin\rho ~ e^{-it \pm i \varphi} \left( [D^{(0)}, \CV(x,t)] + \CV(x,t)\right).
\ee
Comparing the above expression with  Eqs.~(\ref{eq:expandedcommute}),\footnote{Eqs.~(\ref{eq:expandedcommute}) are used along with
$[ K_\pm^{(0)}, \frac{V}{2} ] =  [ \bar{L}_0^{(0)}, K_\pm^{(1)}]$ from the
fact that left- and right-moving sectors commute.} we can identify $K^{(1)}$ as
\be
K^{(1)}_\pm &=&  \int d^2x\sqrt{-g} \sin\rho ~ e^{-it \pm i \varphi}  ~\CV(x,t).
\ee
This operator clearly has non-singular matrix elements between states.  With this identification of $K^{(1)}$, we get
the proper conformal algebra at $O(V^2)$ only if in addition we impose the requirement that $[K^{(1)}_\pm, V]=0$.
For a generic interaction, this is possible only if $[\CV(x,t),\CV(y,t)]=0$.  A coordinate-invariant version of this condition
is that whenever one can chosoe a space-like surface containing the two points $(x,x_0)$ and $(y,y_0)$, that $[\CV(x,x_0),\CV(y,y_0)]=0$.

This discussion makes it clear that any local interaction terms, constructed from AdS fields obeying canonical
commutation relations, will lead to a sensible conformally-invariant theory.  Unitarity then places additional constraints
on these local interaction terms.  In particular, if we require perturbative unitarity for all operator dimensions
$\Delta < \Delta_{\rm Heavy}$, then as discussed in section~\ref{sec:unitarity}, local non-renormalizable interactions
must be suppressed by powers of $1/\Delta_{\rm Heavy}$.  In order to understand this matching in more detail, we now
turn to developing the tools needed to efficiently calculate the CFT anomalous dimensions induced by various local bulk
interactions.

\sect{Dilatation Matrix Elements in Low-energy ECT}
\label{sec:dilatationmatrix}

\subsection{Primary Wavefunctions}

At leading order in perturbation theory, corrections to anomalous dimensions are matrix elements of $V$ between primary states.  In many cases of interest, the building blocks of these matrix elements are amplitudes $\<0|\Phi(x)|\psi\>$ for a bulk operator $\Phi(x)$ to annihilate a primary state $|\psi\>$.  For example, in computing the anomalous dimensions of the two-particle primary states $|n,0\>_2$ in $\f^4$-theory, we must evaluate ${}_2\<n,0|\f^4(x)|n,0\>_2=6|\<0|\f^2(x)|n,0\>_2|^2$.  These ``primary wavefunctions" are highly constrained by symmetry, and we can often compute them very efficiently.  In this section, we 
will discuss their general form, and in the next section we will show how to determine their normalizations.

Scalar primary wavefunctions in  AdS$_{d+1}$ are extremely simple.  Note first that any function annihilated by all the $K_\mu$ must be of the form $f(e^{it}\cos\rho)$.  This is clearest in the embedding space construction, where $(e^{it}\cos\rho)^{-1}=X_{0}-i X_{d+1}$, which is the only linear combination of $X$'s that is killed by all the rotation generators $K_\mu=J_{\mu, d+1} + i J_{\mu,0 }$.  Thus, for scalar $\Phi(x)$,
a primary wavefunction for a state $|\psi \>$ 
with definite energy $\w$ is proportional to
\be
\label{scalarprimarywf}
\<0|\Phi(x)|\psi\> &\propto& \p{e^{it}\cos\rho}^\w,
\ee
where the constant of proportionality vanishes if $|\psi\>$ has nonzero spin.
Related arguments were used in~\cite{andyjuan, andyweiwei}.

More generally, we might be interested in the wavefunction of a tensor operator $\Phi^{a_1\dots a_n}(x)$ in a primary state $|\psi_{\mu_1\dots\mu_l}\>$ with energy $\w$ and spin $l$.\footnote{We use Roman indices $a,b,c,\dots=1,\dots,d+1$ for the tangent space in global AdS$_{d+1}$, and Greek indices $\mu,\nu,\dots=1,\dots,d$ for the Euclidean coordinates of the embedding space. In particular,
 $g_{\mu\nu}=\delta_{\mu\nu}$.   Here, we are writing an element of the spin-$l$ representation of $\SO(d)$ as a traceless symmetric tensor with $l$  $\mu$-indices.}  To determine its general form, we can start by writing down a basis of tensor fields in AdS$_{d+1}$ that are invariant under the action of $K_\mu$.  Since special conformal transformations commute, the associated vector fields $\xi_\mu^a\equiv (K_\mu)^a$ are trivially invariant under Lie derivatives $\cL_{K_\nu}$.  Together with $\z^a\equiv \ptl^a(e^{it}\cos\rho)^{-1}$, they form a $K_\mu$-invariant basis for the tangent space at each point in AdS$_{d+1}$.\footnote{
We could have chosen $\zeta^a$ to be a derivative of any function
of $e^{i t} \cos \rho$, since the $K_\mu$'s would annihilate it.
The choice $(e^{it} \cos \rho)^{-1}$ is convenient since then $\zeta^a$ and
$\xi_\mu^a$ have the same scaling dimension. }  A general primary tensor is therefore just a product of $\xi_\mu$'s and $\z$'s, times a function $f(e^{it}\cos\rho)$.  Note further that
\be
h^{ab}&=&(e^{it}\cos\rho)^2(\xi_\mu^a \xi^{\mu b}+\z^a\z^b),
\ee
where $h^{ab}$ is the metric on AdS$_{d+1}$, so we can trade traces $g^{\mu\nu}\xi_\mu^a\xi_\nu^b$ for factors of $\z^a\z^b$ and $h^{ab}$.  Finally since $\xi_\mu$ and $\z$ are lowering operators for the dilatation generator $D$, a basis for wavefunctions of states $|\psi_{\mu_1\dots\mu_l}\>$ with definite energy $\w$ and spin $l$ is given by
\be
\<0|\Phi^{a_1\dots b_1\dots }(x)|\psi_{\mu_1\dots\mu_l}\> &\sim& (e^{it}\cos\rho)^{\w+n+l}\,\z^{a_1}\cdots \z^{a_n}\p{\xi_{(\mu_1}^{b_1}\cdots\xi_{\mu_l)}^{b_l}-\mbox{traces with $g^{\mu\nu}$}}
\label{eq:tensorprimarybasis}
\ee
(up to possible factors of $h^{ab}$).  Here, the states $| \psi_{\mu_1 \dots
\mu_l} \>$ have been labeled so that their wavefunctions are grouped together
into tensors like the right-hand side of Eq.~(\ref{eq:tensorprimarybasis}),
but one is usually interested in states with definite angular quantum
numbers.  One can obtain the wavefunction for such a state by
projecting the above wavefunctions onto the appropriate polarization.
For instance, in AdS$_4$ we obtain the unique $l=m=2$ two-index wavefunction
by projecting Eq.~(\ref{eq:tensorprimarybasis}) onto the polarization tensor $\epsilon^{(2,2)}_{\mu\nu}$:
\be
\< 0 | \Phi^{b_1 b_2}(x) | 2, 2\> &\propto& (e^{i t} \cos \rho )^{\omega + 2}
(\xi^{b_1}_{(\mu } \xi^{b_2}_{\nu)} - \frac{1}{3} \xi_{\sigma}^{b_1} \xi^{\sigma b_2} g_{\mu\nu} 
) \epsilon^{(2,2)\mu\nu} , \ \ \
\epsilon^{(2,2)}_{\mu\nu} = \left( \begin{array}{ccc}
1 & i & 0 \\
i & -1 & 0 \\
0 & 0 & 0 \end{array} \right) .
\ee

In AdS$_3$, this basis simplifies slightly.  In light-cone coordinates on the boundary, the special conformal generators $\xi_{\pm}^a$ are given in Eq.~(\ref{2dgen}), and we have $g^{\mu\nu}\xi_\mu^a\xi_\nu^{b}=\xi_+^{(a}\xi_-^{b)}$.  Thus, a basis for tensor wavefunctions is given by
\be
(e^{it}\cos\rho)^{\w+n+l} \xi_{+}^{a_1}\cdots \xi_{+}^{a_l}\z^{b_1}\cdots\z^{b_n}, \qquad (l>0)\nn\\
(e^{it}\cos\rho)^{\w+n-l} \xi_{-}^{a_1}\cdots \xi_{-}^{a_{-l}}\z^{b_1}\cdots\z^{b_n}, \qquad (l<0)
\ee
(up to possible factors of $h^{ab}$).  For example, to write the two-index spin-$2$ wavefunction in AdS$_3$, we can use $(\xi_\pm)_a=- i\frac{\sin\rho}{\cos^2\rho}e^{-it\pm i\varphi}(1,\pm1,-i\cot\rho)$ (in coordinates $t,\varphi,\rho$), and find
\be
\<0|\Phi_{ab}(x)|\pm 2\> &\propto& (e^{it}\cos\rho)^{\w+2}(\xi_\pm)_a(\xi_\pm)_b\nn\\
 &\propto&e^{i\w t\pm 2 i\varphi}\cos^\w\rho\tan^2\rho\left(
 \begin{array}{ccc}
 1 & \pm 1 & -i\cot \rho\\
 \pm 1 & 1 & \mp i\cot \rho\\
 -i\cot\rho & \mp i \cot \rho & -\cot^2\rho
 \end{array}
 \right).
 \label{spin2fcn}
\ee

\subsection{Normalization of Primary Two-particle Wavefunctions}

We can extract normalizations of primary wavefunctions by a procedure analogous to the conformal block decomposition of CFT correlators.  Consider the contribution of a scalar primary state $|\psi\>$ of dimension $\w$ and its descendants to the two-point function of a bulk scalar operator $\Phi(x)$,
\be
\sum_{\a=\psi\mathrm{,\ desc.}}
\<0|\Phi(x)|\a\>\<\a|\Phi(x')|0\>.
\label{eq:modesumoverdescendants}
\ee
We know $\<0|\Phi(x)|\psi\>$ is determined by symmetry.  In particular, up to a normalization factor it is the same as the primary wavefunction of a free field, 
\be
\<0|\Phi(x)|\psi\> &=& \frac{1}{N^{\Phi}_\psi}(e^{it}\cos\rho)^\w\nn\\
&=& \frac{N_{\w,0,0}}{N^{\Phi}_\psi}\vol(S^{d-1})^{1/2}\phi_{00}(x) ,
\ee
where $N_{\w,0,0}$ is given in Eq.~(\ref{eq:singleparticlenorm}).  But descendant wavefunctions are determined by the primary wavefunction, so all the $\<0|\Phi(x)|\a\>$ are also proportional to wavefunctions of a free field, given in Eq.~(\ref{eq:singleparticlewavefunctions}) with $\De\to\w$.  Note that $\Phi(x)$ itself need not be a free field, and $|\psi\>$ need not be a single-particle state --- conformal symmetry determines everything up to normalization.  Consequently the sum in Eq.~(\ref{eq:modesumoverdescendants}) is precisely the same as the sum over modes in a free-field two-point function, and the answer is simply a constant times the bulk propagator,
\be
\sum_{\a=\psi\mathrm{,\ desc.}}
\<0|\Phi(x)|\a\>\<\a|\Phi(x')|0\>
&=&
 \frac{N_{\w,0,0}^2}{(N^{\Phi}_\psi)^{2}}\vol(S^{d-1}) K_B(x,x') \nn\\
 &\equiv& \frac{G_\w(z)}{(N_\psi^\Phi)^{2}} ,
\ee
where
\be
G_\w(z) &=& z^{\w/2} F\p{\w,\frac d 2,\w+1-\frac d 2,z}
\ee
and $z=e^{-2\s(x,x')}$, with $\s(x,x')$ the geodesic distance between $x$ and $x'$.  Summing over primary states $|\psi\>$, we find
\be
\<0|\Phi(x)\Phi(x')|0\> &=& \sum_{\psi\mathrm{\ primary}} \frac{G_\w(z)}{(N_\psi^\Phi)^{2}},
\ee
so we can extract the normalizations $N_\psi^\Phi$ by decomposing $\<0|\Phi(x)\Phi(x')|0\>$ into bulk propagators.  To do this in practice, it is useful to exploit the Klein-Gordon equation for the propagator as a function of $z$,
\be
\frac{z^{d/2+1}}{(1-z)^d} \frac{d}{dz}\p{\frac{(1-z)^d}{ z^{d/2-1}} \frac{d}{dz} G_\w(z)} &=& \frac 1 4\w(\w-d)G_\w(z).
\ee
This implies the orthogonality relation,
\be
\oint \frac{dz}{2\pi i}\frac{(1-z)^d}{z^{1+d/2}} G_{d-\a}(z) G_\b(z) &=& \de_{\a\b},
\ee
where the right-hand side uses the fact that the $G_\omega(z)$ are already
normalized with respect to this inner product.
As an example that will be relevant shortly, let us find the normalization of the wavefunction of $\f^2(x)$ in the scalar two-particle primary state $|n,0\>_2$ of dimension $2\De+2n$.  The two-point function $\<0|\f^2(x)\f^2(x')|0\>$ is easily computed from Wick contractions:
\be
\<0|\f^2(x)\f^2(x')|0\>\ \ =\ \ 2K_B(x,x')^2\ \ =\ \ \frac{2}{N_{\De,0,0}^4\vol(S^{d-1})^2}G_\De(z)^2 .
\ee
Applying our orthogonality relation, we get
\be
\frac{1}{(N^{\f^2}_{n,0})^2} &=& \frac {2} {N_{\De,0,0}^4\vol(S^{d-1})^2} \oint \frac{dz}{2\pi i}\frac{(1-z)^d}{z^{1+d/2}} G_\De(z)^2 G_{d-(2\De+2n)}(z)\nn\\
&=& \frac {\G(n+\frac{d}{2})\G(\De+n)^2  \G(2\De+n-\frac{d}{2}) \G(2\De+2n-d+1) }{2 \pi^d n! \G(\frac{d}{2}) \G(\De+n-\frac{d-2}{2})^2 \G(2\De+n-d+1) \G(2\De+2n-\frac{d}{2})} .
\label{eq:wavefnnorm}
\ee
Though we have given the general answer, the above integral tends to be particularly simple in even dimensions where $G_{\De}(z)$ is an elementary function.  For instance, in $d=2$, we have $G_\De(z)=z^{\De/2}(1-z)^{-1}$, and the contour integral essentially just computes coefficients in the Taylor expansion of $(1-z)^{-1}$ around $z=0$.  For use in later sections, let us quote the result in $d=2$ and $d=4$:
\begin{align}
\label{AdS3twoparticlenormalized}
\<0|\f^2(x)|n,0\>_2 &= \frac{1}{\sqrt{2}\pi}(e^{it}\cos\rho)^{2\De+2n} & (d=2) ,\\
\<0|\f^2(x)|n,0\>_2 &= \frac {(\De+n-1)} {\sqrt{2}\pi^2}\sqrt{\frac{(n+1)(2\De+n-3)}{2\De+2n-3}}(e^{it}\cos\rho)^{2\De+2n} & (d=4) .
\end{align}

\subsection{Example Calculation of $V_{nm}$}

We are now in a position to easily calculate the matrix elements
of $V$ for various local AdS bulk interactions.  Let us begin with
the simplest example, which is a quartic interaction in AdS$_3$,
\be
V &=& \frac{\mu}{4!} \int d^2 x \sqrt{-g} \phi^4(x) .  
\ee
We are specifically interested in the matrix elements
\be
V_{n m} &=& \frac{\mu}{4!} {}_2\< n, 0 | \int d^2 x \sqrt{-g}
\phi^4(x) | m,0 \>_2 \nn\\
  &=& \frac{\mu}{4!} \int d^2 x \sqrt{-g}
{}_2\< n, 0 |  : \left( \sum_{n,l} \phi_{nl}(x)
a_{nl} + \phi^*_{nl}(x) a^\dagger_{nl} \right)^4 : | m,0 \>_2,
\ee
where $: (\dots) :$ denotes normal ordering, which we will not write explicitly
from now on.  There are $4!$ possible contractions of the external states, each of which
gives the same contribution, summing to
\be
V_{n m } &=& \frac{\mu}{4} \int_{0}^{2\pi} d \varphi \int_{0}^{\pi/2} d \rho
\frac{\sin \rho}{\cos^3 \rho}  {}_2\< n, 0 | \phi^2(x) | 0 \>
\< 0 | \phi^2(x) | m , 0 \>_2 .
\ee
Now we can apply the results of the previous two subsections, namely that 
the wavefunctions $\< 0 | \phi^2(x) | n, 0 \>_2$ are completely determined
by conformal symmetry!  Plugging in (\ref{AdS3twoparticlenormalized}), we can trivially perform the 
integration above to obtain
\be
V_{nm} &=& \frac{\mu }{8\pi (2\Delta + n+m  -1)} .
\ee
Of course, the anomalous dimension $\gamma(n,0)$ of $| n, 0 \>_2$ is just
$V_{nn}$, so we have
\be
\gamma(n,0) &=& \frac{\mu }{8\pi(2\Delta + 2n -1)},
\ee
which reproduces the result in \cite{polchinski} based on analysis of the
four-point AdS boundary correlator.
Note that this provides a simple example of why in section~\ref{sec:unitarity} we could take
$V_{n,n+1} \approx \gamma(n,0)$ at large $n$ -- the wavefunctions
for $|n, 0 \>_2$ and $|n+1,0 \>_2$ are negligibly different at large $n$,
so the matrix element of $V$ between them is nearly the same as the matrix
element between $|n, 0 \>_2$ and itself.

Let us pause to emphasize the simplicity of this calculation. 
The integrations we had to do above were extremely simple.  Even the machinery developed
in the previous sections, which was designed solely to construct the two-particle
wavefunctions and was not specific to any individual
AdS bulk interaction, required little calculation. The form of the wavefunctions
followed very simply from the property of the states being primary and
scalar, and their normalization followed essentially from expanding
$(1-z)^{-1}$ around $z=0$.   Nowhere did we have to calculate a
four-point boundary correlation function in AdS, or to extract log terms.  It is
also completely manifest that no primary state with spin $l>0$ can
get a contribution from $\phi^4(x)$; there simply is no spin-$l$
primary wavefunction that can be constructed without AdS-Lorentz indices
unless $l=0$.  

By projecting onto the double-trace primary states at
the very beginning of the calculation, rather than near the end,
one can circumvent having to deal with significantly more complicated
structures which are not particularly relevant to the calculation of anomalous dimensions. 
This should make it clear that the present approach is capable of greatly simplifying the
analysis of the behavior of anomalous dimensions in the $1/N$ expansion.
In particular, we will now turn to a discussion of the scaling behavior
of $\gamma(n,l)$ for various AdS interactions.  We will see that dimensionless 
quantities like $n$ and $\Delta$ can in fact be interpreted
as dimensionful quantities when they are large (compared to 1),
and that they obey their own rules of dimensional analysis.

\subsection{Dimensional Analysis with $n$}

The interaction $\phi^4$ in AdS$_3$ we considered in the previous section
was renormalizable, i.e. $\mu$ had mass-dimension 1, and the anomalous
dimension $\gamma(n,0)$ decreased like $\sim n^{-1}$ at large $n$.
This suggests that we should assign mass-dimension zero to $\gamma(n,0)$
and mass-dimension 1 to $n$, so that at large $n$ dimensional analysis
forces the correct $n$-dependence $\gamma(n,0) \sim \mu/n$. 
 How does this work for other examples,
in particular non-renormalizable operators?  Consider the first few
non-renormalizable four-point interactions in AdS$_3$:
$\mu^{-1} \phi^2 (\nabla \phi)^2$, $\mu^{-3} (\nabla \phi)^4, $ and $
\mu^{-5} (\nabla_\mu \nabla_\nu \phi)^2$.
In all these cases, $\gamma(n,l)$ was calculated
in \cite{polchinski} based on four-point correlators; we show in
Appendix \ref{app:dphi4} how to reproduce these results using the
present methods.  The first is accidentally renormalizable, since it
may be reduced to $-\frac{m^2}{3\mu} \phi^4$ by integration by parts and using the
equations of motion.  However, when we calculate $V_{nn}$ from this
operator, its accidental renormalizability arises from a cancellation
among the different contractions of the $\phi$'s, and it is illuminating
to consider them separately,
\be
 {}_2\< n, 0 | (\nabla \phi)^2 \phi^2 | n, 0 \> _2 &=&
  2\,{}_2\< n, 0 | (\nabla \phi)^2 | 0 \> \, \< 0 |
 \phi^2 | n, 0 \>_2 +  \nabla^\mu {}_2\< n, 0 | \phi^2 | 0\>\,
\nabla_\mu \< 0 | \phi^2 | n, 0 \>_2  . \nn\\
\ee
The first of these may easily be evaluated, since $ (\nabla  \phi)^2
 = \half \nabla^2 \phi^2 - \phi \nabla^2 \phi \cong
(\half m_n^2 - m^2) \phi^2$, where $m_n^2 = 4(\Delta+n) (\Delta + n -1)$
is the effective mass of the two-particle primary operator (i.e.,
its scalar wavefunction obeys $(\nabla^2 - m_n^2)\phi^2 = 0$) .  The second term
is only slightly more involved.  In both cases, one can clearly see the
additional powers of $\Delta + n$ being pulled down from the $\d_t $ and
$\d_\rho$ derivatives to make the contribution at large $n$ behave like
$n^2$ times the $\phi^4$ result.  The reduction to a lower-dimensional
operator due to the equations of motion
is specific to $(\nabla \phi)^2 \phi^2$, and
in general additional derivatives behave like additional powers of $n$,
exactly as is necessary for dimensional analysis with $n$'s to work.
It follows that any four-point interaction in AdS$_3$ with dimension
$p$ leads to growth in $\gamma(n,0)$ like $\sim n^{p-3}$.

We can generalize these results to a quartic $\phi$ interaction in any
dimension, using our previous results for the scalar two-particle wavefunctions. 
To consider the large $n$ behavior arising from an arbitrary quartic interaction, it suffices to 
calculate the scaling of $\gamma(n,0)$ for
$\phi^4$, since as we have seen above, additional derivatives in the
interaction just pull down more powers of $\Delta + n$.  More concretely,
if we consider a quartic interaction in AdS$_{d+1}$
\be
V &=& \frac{\mu^{3-d}}{4!} \int d^d x \sqrt{-g} \phi^4(x) ,  
\ee
using the general 2-particle wavefunctions we can readily calculate
\be\label{eq:gammagenerald}
\gamma(n,0) &=& \frac{\mu^{3-d}}{4} \int d\Omega \int_{0}^{\pi/2} d \rho
\frac{\sin^{d-1} \rho}{\cos^{d+1} \rho}  {}_2\< n, 0 | \phi^2(x) | 0 \>
\< 0 | \phi^2(x) | n , 0 \>_2 \nn\\
&=& \frac{\mu^{3-d} \pi^{d/2}}{4 (N_{n,0}^{\phi^2})^2}    \frac{\Gamma(2\Delta + 2n -\frac{d}{2})}
{ \Gamma(2\Delta + 2n)} .
\ee
Now from Eq.~(\ref{eq:wavefnnorm}), we can read off that the
wavefunction coefficient-squared $(N_{n,0}^{\phi^2})^{-2}$ grows like
$[n(n+\Delta)(n+2\Delta)]^{(d-2)/2}$ at large $n$, whereas the ratio of gamma functions in Eq.~(\ref{eq:gammagenerald})
scales like  $(n+\Delta)^{-d/2}$.
Consequently,  we have that $\gamma(n,0)$ for $\phi^4$ at
large $n$ scales like
\be
\gamma(n,0) &\sim& \mu^{3-d} \frac{[n(n+2\Delta)]^{(d-2)/2}}{\Delta+n} \stackrel{n\gg \Delta}{\longrightarrow} \left(\frac{\mu}{n}\right)^{3-d},
\label{eq:phi4anydim}
\ee
which verifies explicitly that dimensional analysis works with any quartic
scalar contact interaction in any dimension.  Note that we could have easily predicted this
behavior simply by demanding that $\gamma(n,0)$ is proportional to the ``dimensionless" combination $(\mu/n)^{3-d}$ built out of
the ``dimensionful" parameters $\mu$ and $n$, since the $\mu$ scaling is just determined by the dimension of the interaction.
Roughly speaking, $\Delta + n$ is an ``energy'' and $[n(n+2\Delta)]^{1/2}$ is a ``momentum'', and the scaling simplifies
when $n \gg \Delta$ because energy and momentum become the same in this ``relativistic'' limit. We will 
discuss this connection in detail in section \ref{sec:flatspace}.

\sect{Heavy field Exchange}
\label{sec:heavyexchange}

Finally we will turn to the exchange of a heavy scalar in AdS, which will help to illustrate
the real power of the techniques developed in the previous sections and will let us further explore
the meaning of AdS effective field theory in terms of CFT quantities.  Heavy scalar exchange
contributions to CFT four-point functions have been studied using a variety of
techniques (see e.g.~\cite{Liu:1998th,D'Hoker:1998mz,D'Hoker:1999ni,Hoffmann:2000tr,Hoffmann:2000mx,Dolan:2000ut,Cornalba:2006xm,Cornalba:2007zb}), 
but extracting information about anomalous dimensions has proven to be relatively difficult using the standard methods.  
Here we will see that the formalism developed above is well suited to studying this problem.

To be concrete, we will consider the bulk interaction
\be\label{eq:Vfull}
V &=& \frac{\mu^{\frac{5-d}{2}}}{2} \int d^d x \sqrt{-g} \phi^2(x) \chi(x)
\ee
between massive scalars $\phi(x)$ and $\chi(x)$ in AdS$_{d+1}$.  We will focus on
the case of $d < 6$ so that this interaction is a renormalizable operator.  In the limit
that $m_{\chi} \gg m_{\phi}$ we can integrate out $\chi$ and obtain an effective
field theory with contact terms
\be\label{eq:Veff}
V_{eff} &\sim& \frac{\mu^{5-d}}{m_{\chi}^2} \int d^d x \sqrt{-g} \phi(x)^4 + \dots
\ee

Below we will compare the contributions to the anomalous dimensions of
the $\phi$ double-trace operators from the full interaction Eq.~(\ref{eq:Vfull}) to the
contributions from the effective field theory truncation Eq.~(\ref{eq:Veff}).  We will find that the
effective Lagrangian indeed approximates the full result when
$n \ll \Delta_{\chi}$, but deviates from it when $n \sim \Delta_{\chi}$, eventually
growing and violating the unitarity constraint discussed in section~\ref{sec:unitarity}.  In the full
theory this growth is cut off by considering more and more terms in the effective
Lagrangian, and in the CFT this amounts to ``integrating in" the operator
sourced by $\chi$.  In fact, as we will see shortly, one can even observe the appearance of a
resonance in $\gamma(n,0)$ near $n \sim \Delta_{\chi}$, completely analogous
to the resonance that appears in scattering amplitudes!  We will have more to
say about this below, but it should be clear that much of the intuition gained from
thinking about effective field theories can be directly carried over to effective
conformal theories.

\subsection{S-channel Scalar Exchange}
\label{sec:schannel}

In order to simplify the problem we will start by focusing on scalar exchange
in the s-channel, which only contributes to the $l=0$ anomalous dimensions
$\gamma(n,0)$.  Since it is straightforward to identify the s-channel contractions
of the quartic operators in the low-energy theory, we will be able to compare the
full s-channel scalar exchange contribution at all energies to the low-energy effective
theory.

Now let us compute the corrections to the anomalous dimensions $\gamma(n,0)$
using old-fashioned perturbation theory.  Since scalar exchange requires two insertions
of the interaction in Eq.~(\ref{eq:Vfull}) we must go to second order in perturbation theory.
The anomalous dimensions are then given by
\be
\gamma(n,0) &=& \sum_{\alpha} \frac{|\<\alpha| V |n,0\>_2|^2}{E_{n} - E_{\alpha}} ,
\ee
where $E_{n} \equiv E_{n,0} = 2\Delta+2n$ and $\alpha$ runs over all states with one $\chi$ particle and either zero, two, or four $\phi$ particles.

\begin{figure}[t!]
\begin{center}
\includegraphics[width=0.85\textwidth]{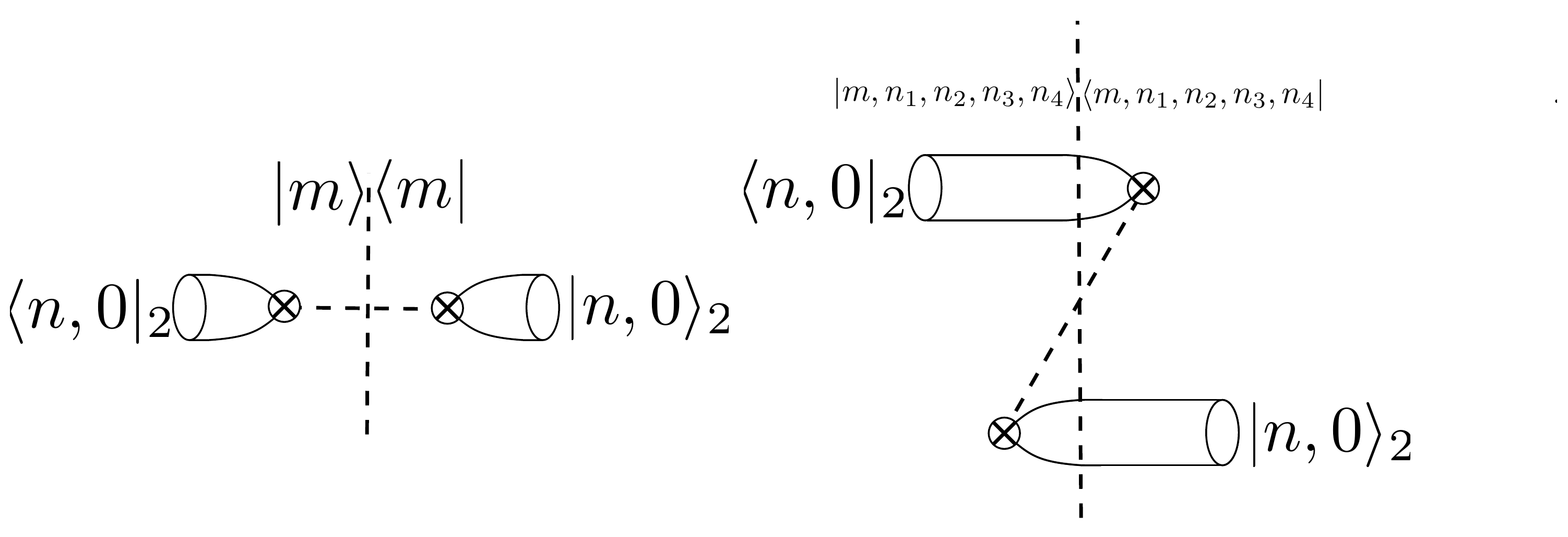}
\caption{One-particle (left) and five-particle (right) intermediate
state diagrams contributing to the ``s-channel''.
\label{fig:schanneldiagrams}}
\end{center}
\end{figure}

S-channel exchange corresponds to intermediate states with
one $\chi$ particle as well as the ``time reversed" intermediate states with
four $\phi$ particles and one $\chi$ particle (see Fig.~\ref{fig:schanneldiagrams}). Since time reversal is equivalent to taking
$E_{n} \rightarrow -E_{n}$, the full s-channel contribution is
given by a sum over one-particle states
\be\label{schannelgamma}
\gamma(n,0) &=& \sum_{m=0}^{\infty} |\<\chi; m,0| V |n,0\>_2|^2\left(\frac{2 E^{\chi}_m}{E_{n}^2 - E^{\chi 2}_{m}}\right)
\ee
where $E^{\chi}_m = \Delta_{\chi} + 2m$, and we have used the fact that angular momentum conservation allows only $l=0$ states to contribute.

Now we can easily calculate the needed matrix element using the explicit form of the one-particle and two-particle states obtained in the previous sections
\be
\<\chi; m,0| V |n,0\>_2 &=& \frac{\mu^{\frac{5-d}{2}}}{2} \int d^d x \sqrt{-g} \<\chi; m,0| \chi(x) | 0 \> \< 0 | \phi^2(x) | n,0 \>_2 \nn\\
 &=&  \frac{\mu^{\frac{5-d}{2}}}{2 \sqrt{\textrm{vol}(S^{d-1})} N_{m,0}^{\chi}  N_{n,0}^{\phi^2}}  \int d\Omega \int_0^{\pi/2} d\rho \frac{\sin^{d-1} \rho}{\cos^{d+1} \rho} \cos^{E_n+\Delta_{\chi}} \rho F(-m,\Delta_{\chi}+m,\frac{d}{2},\sin^2\rho) \nn\\
&=& \frac{\mu^{\frac{5-d}{2}}}{N_{n,0}^{\phi^2}}  \sqrt{\frac{\pi^{d/2} \Gamma(\frac{d}{2}+m)  \Gamma(\Delta_{\chi}+m)}{8 \Gamma(\frac{d}{2}) m! \Gamma(\Delta_{\chi}-\frac{d-2}{2}+m)}}    \frac{ \Gamma(\frac{\Delta_{\chi}+E_n-d}{2}) \Gamma(\frac{E_m^{\chi}-E_n+2}{2})}{\Gamma(\frac{\Delta_{\chi}-E_n+2}{2}) \Gamma(\frac{E_m^{\chi}+E_n}{2}) } .
\label{twoparticleoneparticleoverlap}
\ee

Finally, we can square this and perform the sum over $m$ in Eq.~(\ref{schannelgamma}), which for general $d$ may be written in terms of $_4F_3$ hypergeometric functions
\be
\gamma(n,0) 
&=& -\frac{\mu^{5-d} \pi^{d/2}  }{8 (N_{n,0}^{\phi^2})^2} \frac{\Gamma(\Delta_{\chi})\Gamma(\frac{\Delta_{\chi}+E_n-d}{2})^2}{\Gamma(\Delta_{\chi}-\frac{d-2}{2})\Gamma(\frac{\Delta_{\chi}+E_n}{2})^2} \\
&&\times \left[\frac{{}_4F_3\left(\left\{\frac{\Delta_{\chi}-E_n}{2},\frac{\Delta_{\chi}-E_n+2}{2},\Delta_{\chi},\frac{d}{2}\right\},\left\{\frac{\Delta_{\chi}+E_n}{2},\frac{\Delta_{\chi}+E_n}{2},\Delta_{\chi}-\frac{d-2}{2}\right\} , 1 \right)}{\Delta_{\chi}-E_n} \right. \nn \\
&& \qquad \left. + \frac{{}_4F_3\left(\left\{\frac{\Delta_{\chi}-E_n+2}{2},\frac{\Delta_{\chi}-E_n+2}{2},\Delta_{\chi},\frac{d}{2}\right\},\left\{\frac{\Delta_{\chi}+E_n}{2},\frac{\Delta_{\chi}+E_n+2}{2},\Delta_{\chi}-\frac{d-2}{2}\right\} , 1 \right)}{\Delta_{\chi}+E_n} \right] \nn
\ee

\begin{figure}[t!]
\begin{center}
\includegraphics[width=0.75\textwidth]{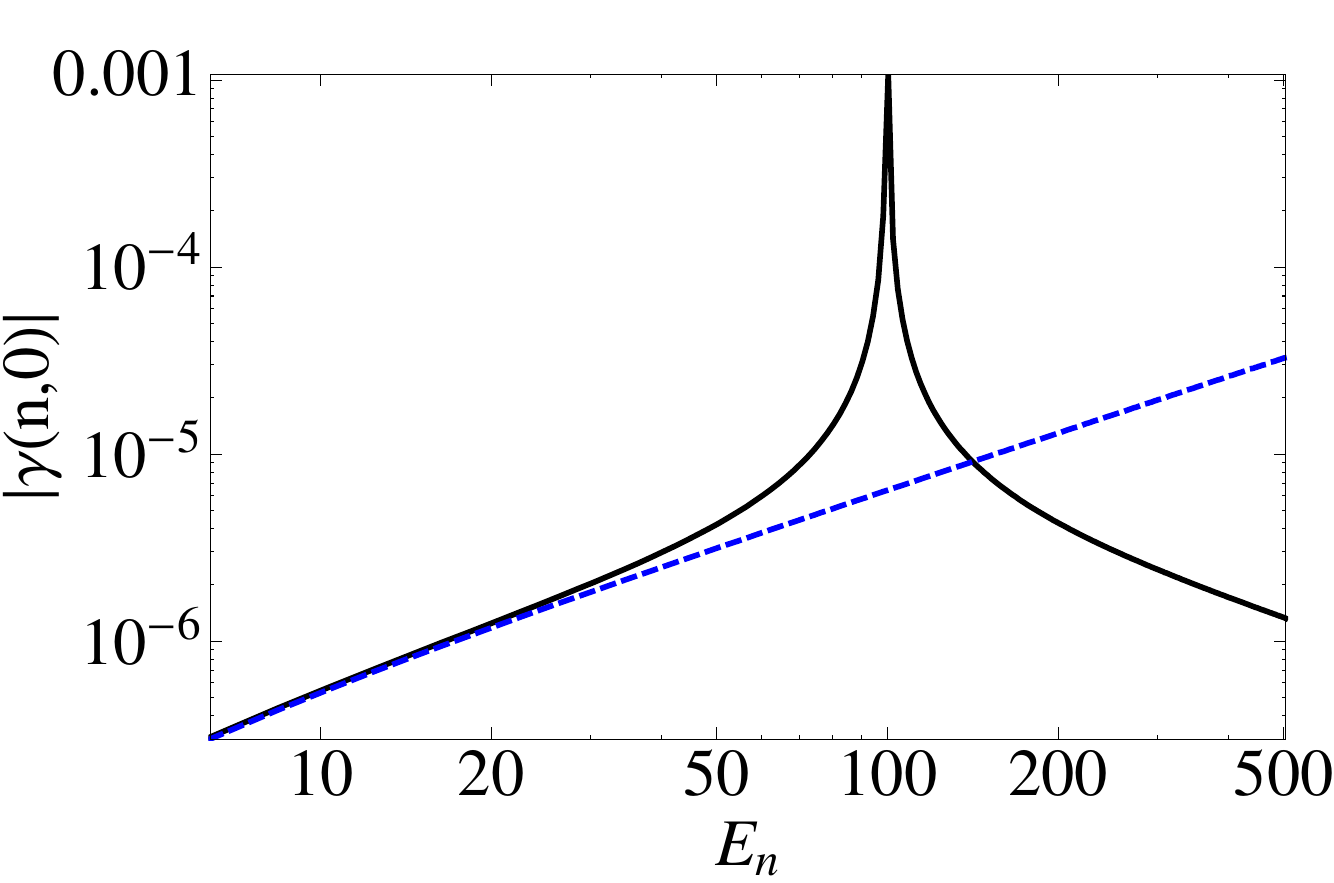}
\caption{Plotted are the contributions to $|\gamma(n,0)|$ from s-channel scalar exchange (solid line) and s-channel contractions of the low-energy $\phi^4$ interaction (dashed line) in AdS$_5$ with $\Delta_{\chi}=100.1$ and $\Delta=2.2$.
\label{fig:scalarexchangedims} }
\end{center}
\end{figure}

It is easy to see that this expression has a pole at $E_n = \Delta_{\chi}$, and close to this value there is a resonance-like
enhancement of $\gamma(n,0)$.  We can clearly see this behavior in Fig.~\ref{fig:scalarexchangedims}, where
we have specialized to AdS$_5$ and chosen $\Delta = 2.2$ and $\Delta_\chi = 100.1$ for illustrative purposes.  Actually,
while the expression we derived blows up at $E_n = \Delta_{\chi}$, if we were to go to higher order in perturbation theory we
would see that the resonance gets smoothed out and has a finite width $\Gamma \sim \sum |\<\chi|V|\phi^2\>|^2$, corresponding
to the fact that $\chi$ has a finite lifetime in AdS due to the trilinear interaction.

At large $n$ we see that $\gamma(n,0)$ has a $1/n$ falloff in AdS$_5$, and more generally the large $n$ behavior scales like $1/n^{5-d}$.
This is precisely what we would expect based on our ``dimensional analysis" discussion in the previous section, since $\gamma(n,0)$ should be proportional
to the ``dimensionless'' combination $(\mu/n)^{5-d}$.

\subsection{Matching Between Low and High Energies}

On the other hand, at small $n$ there is another ``scale" in the problem (namely $\Delta_{\chi}$), and the
behavior is dominated by the bulk contact interactions in the effective field theory suppressed by this scale.  We can concretely see
this behavior in Fig.~\ref{fig:scalarexchangedims}, where we have in addition plotted the
contribution to $\gamma(n,0)$ from the s-channel contractions of the low-energy bulk contact interaction term $\phi^4$.  At smaller values of $n$,
both functions behave roughly like $\sim n$ (as expected from dimensional analysis of the $\phi^4$ interaction), but while the full correction
then passes through a resonance at $E_n = \Delta_{\chi}$ and transitions to its large $n$ behavior, the contribution from the
$\phi^4$ interaction continues to simply rise like $\sim n$.
Because this operator is non-renormalizable,
we see continued growth in $\gamma(n,0)$ as $n$ increases;
however, rather than continuing indefinitely
and violating unitarity, the growth is cut off in the full theory
by ``integrating in'' the heavy primary, exactly as we would expect from effective field
theory in AdS.

To better understand the matching to low energies let us try to analytically extract the leading low-$n$ behavior of
$\gamma(n,0)$ by taking the large $\Delta_{\chi}$ limit.  To do this we can approximate
the $\Gamma$ functions in the sum using the expansion
\be
\frac{\Gamma(z+a)}{\Gamma(z+b)} &=&
   z^{a-b} \left( 1 + \frac{(a+b-1)(a-b)}{2z} + O\left(\frac{1}{z^2} \right)
\right).
\ee
Also in this limit we can take $E_m^{\chi}/(E^{\chi 2}_{m}-E_n^2) \approx 1/E_m^{\chi}$.  Finally, the sum over $m$
can be approximated as an integral in the limit of large $\Delta_\chi$ using an Euler-Maclaurin expansion.
Putting everything together, we have the limiting behavior
\be
\gamma(n,0) &\approx&  -\frac{\mu^{5-d} \pi^{d/2}  }{4 (N_{n,0}^{\phi^2})^2\Delta_{\chi}^{2} } \left(\frac{2^d\Delta_{\chi}^{2E_n-d}}{\Gamma(\frac{d}{2})}   \int_{0}^{\infty} dm \frac{\Gamma(\frac{d}{2}+m)\left(\Delta_{\chi}+m\right)^{d/2-1}}{\Gamma(1+m)\left(\Delta_{\chi} + 2m\right)^{2E_n -1} } + \frac{2^{d-1}}{\Delta_{\chi}^{d/2}} + \dots \right)\nn\\
&\approx& -\frac{\mu^{5-d} \pi^{d/2}  }{4 (N_{n,0}^{\phi^2})^2\Delta_{\chi}^{2} }  \left( \frac{2^d}{\Gamma(\frac{d}{2})}\int_{0}^{\infty} dx \frac{(x(1+x))^{d/2-1}}{(1+2x)^{2E_n-1}}\right) + O\left(\frac{1}{\Delta_{\chi}^3}\right) \nn\\
&\approx& -\frac{\mu^{5-d} \pi^{d/2}}{4 (N_{n,0}^{\phi^2})^2  \Delta_{\chi}^{2} }  \frac{\Gamma(E_n-\frac{d}{2})}{\Gamma(E_n)}+ O\left(\frac{1}{\Delta_{\chi}^3}\right)
\ee
which is precisely the form that we found in Eq.~(\ref{eq:gammagenerald}) corresponding to a $\phi^4$ interaction in AdS$_{d+1}$.

\subsection{T- and U-channels}

The remaining contributions to $V_{nn}$ for scalar exchange come from three-particle intermediate states, where the $\phi^2\chi$ interaction creates a $\chi$ particle and both creates and destroys a $\phi$ particle.  Note that while the $s$-channel contribution may be alternatively written in terms of an integral over the primary wavefunctions of local operators
\be
\gamma_s(n,0) &\propto& \int d^dx\, d^{d+1}x'  \sqrt{-g}  \sqrt{-g'} {}_2\<n,0|\phi^2(x)|0\> K_B^\chi(x,x') \<0|\phi^2(x')|n,0\>_2,
\ee
the $t$- and $u$-channels depend on non-local primary wavefunctions,
\be
\gamma_{t,u}(n,l) &\propto& \int d^dx\, d^{d+1} x' \sqrt{-g} \sqrt{-g'} {}_2\<n,l|\phi(x)\phi(x')| 0\> K_B^\chi(x,x') \<0|\phi(x)\phi(x')|n,l\>_2
\ee
which are not completely fixed by symmetry.  Symmetry does imply, e.g., that
\be
\<0|\phi(x)\phi(x')|n,0\>_2 &\sim& (e^{it}\cos\rho)^{\De+n}(e^{it'}\cos\rho')^{\De+n}f(\s,y),
\ee
where $y=(e^{it}\cos\rho)/(e^{it'}\cos\rho')$ and $\s$ is the geodesic distance between $x$ and $x'$.  We could then use the 
Klein-Gordon equation in $x$ or $x'$ to solve for the function $f$.  However, we will not continue with this analysis in the 
present paper.  The $s$-channel contains most of the interesting physics, including the resonance effect discussed above.  
Further, we will develop a full understanding of all channels at large $n$ (with $\De_\chi, \De$ arbitrary) in the next section.

\sect{Emergence of the Flat-Space S-Matrix from $\gamma(n,l)$}
\label{sec:flatspace}

An important goal of the AdS/CFT correspondence that has been pursued
since its proposal~\cite{Polchinski:1999ry,Susskind:1998vk,Balasubramanian:1999ri,Giddings:1999qu,giddingslocality,Jevicki:2005ms,Gary:2009ae,Gary:2009mi,Okuda:2010ym}
is to learn how information about the S-matrix of the
bulk theory may be extracted from knowledge of the CFT.  This is significantly
more complicated than gaining information in the other direction,
largely because it is difficult to eliminate the boundary effects of
the AdS curvature when the theory being used to probe the S-matrix lives
solely on the boundary.  Various approaches have been taken to
get around this issue, frequently employing the construction in the CFT of
wavepackets designed to collide in the interior of AdS and extract information
about divergences in the resulting interactions.

Here we will take a different approach, based on anomalous dimensions of primary operators, which are more natural quantities from the point of view of the CFT.  We have seen in the preceding sections that $\g(n,l)$ can be computed directly via an AdS scattering process with particular external wavefunctions.  Thus, it is reasonable to expect that we can extract information about the flat-space S-matrix from $\g(n,l)$ in the limit that the energy of this scattering process becomes much larger than the AdS curvature scale.  Remarkably, it turns out this information is encoded very simply.  In the limit $n\gg 1$, two-particle primary states just become flat-space spherical waves with opposite spatial momentum in the frame of the center of AdS.  Consequently, matrix elements ${}_2\<n,l|V|n,l\>_2$ literally become the partial wave expansion of the flat-space S-matrix up to a normalization factor,
\be
\label{eq:flatspacefromgamma}
\CM(s,t,u)^{d+1}_{\textrm{flat space}} &=& \frac{(4\pi)^{d}}{\vol(S^{d-1})}\frac{E_n}{\p{E_n^2-4\De^2}^{\frac{d-2}2}}\sum_l [\g(n,l)]_{n\gg l}\,r_lP^{(d)}_l(\cos\th),
\ee
where the total flat-space energy is $2E = E_n = 2\Delta+2n$ (still in units
of $R=1$), and the Mandelstam variables are defined 
in the usual way with $s=(2E)^2$, $t=-2p^2(1-\cos\th)$, $u=-2p^2(1+\cos\th)$,
and $p^2=E^2-\Delta^2$.
One must formally take the large $n$ limit of $\gamma(n,l)$ before substituting
into the above formula when constructing the flat-space amplitudes.\footnote{
When there are additional CFT parameters such as $\Delta$ (or $\Delta_\chi$
in section \ref{sec:heavyexchange}) that correspond to mass terms,
one must take these to be large as well in order to see their effects
in the scattering amplitude.  More formally, one takes 
$E_n = 2(\Delta+n)k,\, p^2 = n (n+2\Delta)k^2,\, m = \Delta k,\, \dots$ and takes $k\rightarrow 0$ with $E_n,\, p,\, m,\, \dots$
fixed.  There may not always be a free parameter within the CFT that
allows one to take $\Delta$ large; in such cases, Eq.~(\ref{eq:flatspacefromgamma})
 obtains the amplitude $\CM$ with $m=0$.  There is an important caveat here;
the presence of such massless fields in the bulk theory can lead to
infrared-divergent scattering amplitudes, and for such quantities the
left-hand side of
Eq. (\ref{eq:flatspacefromgamma}) would have to be modified to include 
AdS boundary effects.  Thus, if one is not free to dial $m \gg k$ in the
CFT, then one should apply (\ref{eq:flatspacefromgamma}) only to
amplitudes that are infrared-safe in the $m\rightarrow 0$ limit. }
 This correspondence between $\CM(s,t,u)^{d+1}_{\textrm{flat space}}$ and $\g(n,l)$ holds whenever $n$ is much greater than 1.  In particular, it allows us to probe the S-matrix even away from singularities in the four-point function, as was done previously.  
For instance, in section \ref{sec:schannel}, we saw that the anomalous dimensions are sensitive to the behavior of the
S-matrix for scalar exchange at all energies, from far below the
intermediate particle mass, through the resonance, and to far above it.
Singularities in the four-point function from non-renormalizable interactions in the
bulk are unlikely to occur in isolation in an effective AdS theory,
since all non-renormalizable operators tend to become important at
around the same scale.  So we expect it will prove convenient to have a method for extracting S-matrix elements that does not depend on isolating such singularities.

Why should we expect $\gamma(n,l)$ to probe flat space at large $n$?  To a large extent,  it is
because the primary wavefunctions $ \sim \cos^{2\Delta + 2n} \rho$ are 
extremely peaked near $\rho \sim 0$ in this limit.
Since the contribution to $\gamma(n,l)$ is dominated
by the interior of AdS, we expect that the AdS radius
$R$ will become negligible, and the dynamics will be increasingly well
described by flat-space scattering.  More precisely, $\cos^{2\Delta+2n}\rho$
becomes proportional to a delta function at $\cos \rho =1$ as $n$
is taken to $\infty$, so the integral over the bulk may be restricted
to smaller and smaller regions around $\rho=0$.
One may take the large $n$, small
$\rho$ limit by restoring factors of $R$ (as well as $k \equiv 1/R$) 
and taking $R\rightarrow \infty$ with $n/R$ and $r = \rho R$ fixed.  The metric in 
the new coordinates (also making the replacement $t\rightarrow t/R$) is
\be
\label{eq:metricwithunits}
ds^2 &=& \frac{1}{\cos^{2}(r/R)} ( -dt^2 + dr^2 + R^2 \sin^2(r/R) d\Omega^2),
\ee
which approaches the flat-space metric for small $r/R$.
The primary wavefunction then becomes suppressed by an 
exponential damping term $\cos^{2n}\rho \sim e^{- n (k r)^2}$
%%%% This equals:
%$\exp\left[ -\half pr(r/R+ r^3/6R^3 +\dots) \right]$
%%%%
at the scale $r \sim \frac{1}{k\sqrt{n}}$.

However, we can also represent the two-particle primary wavefunctions as a sum
over products of one-particle wavefunctions.  Moreover, deep in the interior
of AdS, it is straightforward to see that the one-particle wavefunctions in Eq.~(\ref{eq:singleparticlewavefunctions}) can 
be approximated by flat-space spherical waves
(see e.g.~\cite{giddingslocality}).   That is, the one-particle wavefunctions become
\be
\phi_{nlJ}(x) &=& \frac{1}{N_{\Delta, n,l}} e^{i E_{n,l} k t} Y_{lJ} (\Omega)
\sin^l (k r) \cos^\Delta (k r) F\left( - n, \Delta+l+n,
l+\frac{d}{2}, \sin^2\left( k r \right) \right)
\nn\\
&\stackrel{k r \ll 1}{\propto}& \frac{1}{(kr)^{\frac{d-2}{2}}} e^{i E_{n,l} k t} Y_{lJ}(\Omega) J_{l+(d-2)/2}(E_{n,l} k r),
\ee
which is a flat-space spherical wave in $d+1$ dimensions with energy $E_{n,l} k$ and angular momentum $l$.
Thus, we expect the two-particle primary wavefunctions in this limit to look like a sum over products of flat-space spherical waves 
(or alternatively plane waves, using the standard decomposition).  

In the next two subsections we will explore more carefully the way in which momentum conservation emerges at large $n$, forcing these 
waves to have opposite spatial momentum so that matrix elements of $V$ look precisely like flat-space scattering amplitudes in the center-of-mass frame.  
We will approach this question from both the CFT and bulk perspectives.  This will eventually lead to a derivation of Eq.~(\ref{eq:flatspacefromgamma}), and 
we will then check it in a number of examples.

\subsection{Emergence of Momentum Conservation}
\label{sec:flatmomcons}
Translation invariance and momentum conservation of amplitudes
must emerge in the flat-space limit.
In particular, one would like to see how delta-functions of the total
momentum emerge in the overlap between two-particle primary states with
one-particle states.  Since a primary state with large $n$ carries
zero momentum (as we will see explicitly in the next subsection), what must 
emerge is something like the flat-space relation $\< P| p_1, p_2 \> \propto \delta(\vec{p}_1 + \vec{p}_2)$ for $\vec{P}=0$,
where $|P\>$ denotes a two-particle state with center-of-mass four-momentum $P$,
and $|p_1,p_2\>$ is a tensor product of one-particle states with four-momenta
$p_1$ and $p_2$.   We can look for this behavior in the explicit form of the overlap of two-particle
primary states with one-particle states.
To begin, let us consider more carefully what these overlaps look like in flat space.
Since we are interested in primary states, we will consider a flat-space two-particle state
$|2E, 0\>_2$ with zero momentum and energy $2E$, which we can decompose into one-particle states 
as
\be
\label{eq:flatspaceoverlap}
| 2E,0 \>_2  &=& \int \frac{d^d p_1}{2 E_1 (2\pi)^d} \frac{d^d p_2}{2 E_2
(2\pi)^d}
\sqrt{\frac{E_1 E_2}{E p_1^{d-2}} }  (2\pi)^{d+1}\delta(2E-E_1 -E_2) \delta^{d}(\vec{p}_1 + \vec{p}_2 )
| E_1, p_1 \> |E_2, p_2\> \nn\\
 &=& \frac{1}{(2\pi)^{d-1}} \frac{p^{\frac{d-2}{2}}}{8E^{\frac{1}{2}}}
\int d\Omega \left| E , p \hat{e}\right.  \> \left| E, -p \hat{e}\right. \>,
\ee
where in the last line,  $p = \sqrt{E^2-m^2}$, and the factor
$\sqrt{\frac{E_1 E_2}{E p_1^{d-2}}}$ is inserted to give the
state the norm ${}_2\< 2E',p' | 2E, 0\>_2 \sim \delta(E-E')\delta^{d}(p')$.
Note that the energies of the one-particle states 
are the same, as a consequence of momentum conservation.
In addition, one can see here a general
factor $p^{\frac{d-2}{2}}/E^{\frac{1}{2}}$ that is responsible for the
$E_n$-dependence of the normalization factor in 
Eq.~(\ref{eq:flatspacefromgamma}).

Now we would like to consider the analogous decomposition in the CFT, where we can
write the double-trace primary states $|n,l\>_2$ in terms of products of single-trace states.
We should be able to see that the overlaps at large $n$ are very narrowly peaked on products of 
single-trace states that have nearly equal weights, just as in Eq.~(\ref{eq:flatspaceoverlap}). 
We can extract this overlap without too much difficulty by considering the two-point and three-point functions in the CFT,
the form of which is fixed up to an overall constant coefficient.  We will show this explicitly in 2d, where 
we will not have to deal with additional angular variables, but the arguments are essentially
the same and can be carried out explicitly in any dimension.

To simplify the discussion even further we will focus our attention on just the left-moving sector.
More precisely, we will consider holomorphic operators $\CO(z)$ that depend
on $z$, but not $\bar{z}$.  Let us take $\CO(z)$ to be a single-trace such operator
 with left-moving weight $h=\Delta/2$.  Also, in analogy with the
double-trace primary operators $\CO_{n,l}(x)$ discussed in the rest of the paper,
let us take $\CO_n(z)$ to be a double-trace left-moving primary with weight
$2h+n$. Then $\CO(z)$ and its descendants are in one-to-one correspondence
with the one-particle states $| h; s \>$, and $\CO_n(z)$ with the primary
state $| 2h +n \>_2$.  We will now proceed to compute the overlap
$\< h; s| \< h; n-s |2h+n \>_2$ in order to compare with our expectations
from flat space.

First we will perform the usual Laurent expansion of the operator $\CO(z)$
in terms of creation and annihilation operators.\footnote{
Technically, this is a Laurent expansion only for even integer $\Delta$,
but for convenience we will abuse terminology somewhat and use this term
for general $\Delta$.}
Taking $z=e^\tau$,
we have
\be
\CO(\tau) &=&\sum_{s=0}^\infty  N_s(h) e^{\tau (h+s)} a^\dagger_s +
N_s(h) e^{-\tau (h+s)} a_s,
\ee
where $a_s^\dagger$ creates the one-particle $s$-th descendant
state $| h; s \>$.  The $N_s(h)$ factors are the Laurent coefficients,
which can easily be extracted from the two-point function:
\be
\< \CO (\tau) \CO(0) \> &=&  \frac{e^{\tau h}}{(e^\tau -1)^{2h}}
   = \sum_s \frac{\Gamma(2h+s)}{\Gamma(2h)s!} e^{-\tau(h+s)} = \sum_s N_s^2(h) e^{-\tau (h+s)}.
\ee
 We can obtain the overlap of $|2h+n \>_2$
with the tensor product of one particle states $|h; m\> | h; n-m\>$
by considering a similar expansion of the three-point function.
To do this, we can first evaluate the correlator $\< \CO(\tau) \CO(0) \CO_n(-T) \>$ with $T\rightarrow \infty$ using the Laurent expansion, which gives
\be
\< \CO(\tau) \CO(0) \CO_n(-T) \> e^{(2h+n)T} 
  &\stackrel{T\rightarrow \infty}{=}& \sum_s N_s(h) N_{n-s}(h) e^{-\tau (h+s)} \< h; s | \< h; n-s|
 2h+n\>_2.
\ee
 Alternatively, we can use the explicit form determined by conformal symmetry:
\be
\< \CO(\tau) \CO(0) \CO_n(-T) \> e^{(2h+n)T}
  &\stackrel{T\rightarrow \infty}{=}& c_n e^{-\tau h}
(1- e^{-\tau})^n
 = c_n \sum_s (-1)^s { n \choose s} e^{-\tau (h+s)}, \ \ \
\ee
 where $c_n$ is the OPE coefficient for $\CO_n$ inside $\CO \times \CO$.
Together, these imply that
\be
\< h; s| \< h; n-s |2h+n \>_2 &=&  { n \choose s} \frac{(-1)^s c_n}{N_s(h)
N_{n-s}(h)}.
\label{eq:two-pcl-one-pcl}
\ee
The right-moving sector essentially just introduces
additional quantum numbers for the states and an additional overlap factor symmetric with the above one.

Now let us return to the issue of momentum conservation.  For large $n$, the overlap factors between two-particle 
primaries and single-particle states are strongly peaked at $s = n/2$, which is exactly where the one-particle
momenta are equal in magnitude, corresponding to the expected delta function
$\delta(\vec{p}_1 + \vec{p}_2)$. In fact, by expanding
$s= n/2+m$ in $m$, one obtains the combinatoric suppression factor ${ n  \choose s} \sim
e^{-m^2/n}$, so that momentum conservation emerges with a fuzziness
proportional to $\sqrt{n}$.\footnote{
In 2d, the Laurent coefficients $N_s(h)$ have trivial $s$-dependence at
large $s$ and may be neglected.  For instance, at $h=\half$, $N_s(h)=1$
identically for any $s$.}  

We can see a similar phenomenon in matrix elements of $V$, which are also expected to conserve momentum at large $n$.
For example, let us consider the matrix elements corresponding to the $\phi^2\chi$ interaction considered in section~\ref{sec:heavyexchange}.  
As $n\to\oo$, the overlap Eq.~(\ref{twoparticleoneparticleoverlap}) (in $d$=2) can be approximated as
\be
\<\chi; m,0| V |n,0\>_2 &\to&\frac{(-1)^m\mu^{3/2} \pi^{1/2}}{n N^{\phi^2}_{n,0} 2 \sqrt{2}}\exp\p{\frac{-m(m+\De_\chi)-\De-\frac{\De_\chi}2 + 1 }{n}},
\ee
which is peaked at $m=0$ (zero $\chi$-momentum), again with fuzziness $\sim\sqrt n$.

Curiously, though we do indeed find momentum conservation at large $n$, we also find violations that {\it grow} with $n$.  This is not a contradiction.  In fact, $\sqrt n$ growth is exactly what is needed for emergence of the flat-space S-matrix.  To see this, let us restore the AdS curvature scale $k$, writing the energy as $E=n k$ and the momentum as $p=m k$.  The typical momentum spread is then
\be
\de p &\sim& \sqrt{k E}.
\ee
At a fixed curvature scale the ``uncertainty'' in momentum
grows with $E$, reflecting the fact that primary wavefunctions become more and more localized in position space,
\be
(\cos k r)^{2E/k} &\sim& e^{-(k E)r^2}.
\ee
However, relative to the scale $E$ of our scattering process, the momentum spread goes to zero at high energies
\be
\frac{\de p}{E} \sim \sqrt{\frac{k}{E}} \to 0,
\ee
so the amplitude is momentum-conserving to leading order in $E$.  In other words, as $n\to \infty$, the primary wavefunctions simultaneously become localized at the center of AdS (and thus insensitive to the global geometry), and approach flat-space momentum eigenstates with translationally-invariant interactions.

\subsection{Two-particle Primaries at Large $n$ in AdS$_{d+1}$}
\label{sec:2particleprimariesind+1}

We have seen how a $\sqrt{n}$ fuzziness in momentum conservation emerges
from the CFT perspective.  Now we will try to see this behavior emerge 
 directly in AdS$_{d+1}$, and solve for the behavior of two-particle primary
wavefunctions at large $R$.  In the coordinates (\ref{eq:metricwithunits}), the AdS isometries (\ref{eq:conformalgenerators}) become
\be
\label{flatspacegeneratorlimit}
K_\mu &=& -R\pdr{}{x^\mu}+ix_\mu\pdr{}t+it\pdr{}{x^\mu}+O(t/R,x/R)\\
P_\mu &=& +R\pdr{}{x^\mu}+ix_\mu\pdr{}t+it\pdr{}{x^\mu}+O(t/R,x/R)
\ee
where $x_\mu=r \Omega_\mu$.  Here we see that at leading order, $K_\mu\sim -R\pdr{}{x_\mu}$ is just the flat-space translation generator, so the leading order condition for a two-particle state to be primary is simply that it have zero total spatial momentum.  Hence, near the center of AdS, if we take 
a two-particle primary wavefunction
 $\< 0 | \phi(x_1) \phi(x_2) |\psi\>_2$ to have definite energy $2E$ and definite momentum $\vec{p}$ in the $x_1$ coordinate,
it should behave like a superposition of plane waves in the center of mass frame
\be
\label{zerothorderplanewaves}
\<0|\f(x_1)\f(x_2)|\psi\>_2 &\sim& e^{iE(t_1+t_2)+ip\.(x_1-x_2)} + O(x/R),
\ee
where $E=E_p\equiv\sqrt{p^2+m^2}$. 

This is almost enough to understand why matrix elements between primaries are so closely related to the flat-space S-matrix.  One might worry that primary states behave less like plane waves away from the center of AdS, and that their matrix elements could be sensitive to these effects.  However, by solving for the two-particle primaries at the next order in $1/R$, we will start to see the position-space localization observed in the previous section, which implies that global geometry becomes irrelevant at high energies.

Let us begin with the zero-th order solution Eq.~(\ref{zerothorderplanewaves}), and allow a small perturbation $q$ around zero total spatial momentum,
\be
\label{nextorderanzatz}
\<0|\f(x_1)\f(x_2)|\psi\>_2 &\sim& \int d^dq\, f(q)e^{iE_{p+q}t_1+iE_{p-q} t_2 + i(p+q)\.x_1-i(p-q)\.x_2}.
\ee
Requiring that this be killed by the $O(R)$ and $O(1)$ terms in Eq.~(\ref{flatspacegeneratorlimit}) then implies
\be
\p{E\pdr{}{q_\mu}+2R q_\mu+O(q/E)}f(q) &=& 0.
\ee
Finally, dropping the $O(q/E)$ terms, this has the solution 
\be
f(q) &=& \frac{1}{(\pi k E /2)^{d/4}}e^{-q^2/kE} ,
\ee
 where the normalization has been chosen so that $\int d^d q\,f(q)^2=1$.  We have thus rederived what we observed in the previous section.  Two-particle primaries at large $n$ approach flat-space plane waves, with opposing momenta peaked at $p\sim \sqrt{E^2-m^2}$, up to an uncertainty $\de p\sim\sqrt{kE}$.

An important point is that this momentum uncertainty only occurs in the center of mass degree of freedom.  Performing the $q$-integration, we see that the wavefunction is proportional to $e^{-kE(x_1+x_2)^2/4}$.  In particular, it is not necessarily suppressed when $x_1\sim -x_2\sim R$.  In this regime, $O(x/R)$ corrections could become important, and to fully understand the wavefunctions we would have to solve for these corrections.  However, for the cases we will be considering, the interactions are either
completely local, or we have the exchange of a massive particle, with mass $M \gg 1/R$.  Therefore, the propagator will suppress the amplitude when $|x_1-x_2| \gg 1/M$.  Thus, the combination of the localization of the center of mass as well as the short
range of propagation ensures that as $E$ becomes large the dominant contribution to the amplitude comes from the
flat region in middle of AdS.  When the angular momentum, $l$, of the state is also large, there is a danger that 
the wave function is no longer fully localized in the center of mass coordinate.  For $l=2$, this lack of localization can 
already be seen explicitly in Eq.~(\ref{spin2fcn}) when $\omega$ is small.  
For the large $l$  cases, we therefore require in addition that $E R \gg l$.    
%However, by additionally restricting to states with angular momentum $l\ll pR$, we can ensure that the wavefunctions are suppressed when  $|x_1-x_2| \sim R$ as well.  Physically, this corresponds to restricting the impact parameter $b \sim l/p $ of the two-particle scattering 
%to be small compared to $R$.  
In terms of CFT quantities this requirement amounts to $n \gg l$, which we assume in following.

Localization near the center of AdS in both $x_1$ and $x_2$ means that when we compute matrix elements, the integrals over spatial slices (coming from our interaction $V=\int d^d x \sqrt{-g}\, \CV(x)$) will always converge before $O(x/R)$ effects become important. 
More precisely, we can split up
the integration over $r$ into three different regions:
flat-space scales $0 < r \lesssim y_f E^{-1}$, large scales
$y_l \sqrt{R E^{-1}} \lesssim r < R \pi/2$ containing the boundary of AdS,
and the remaining intermediate region,
containing the transition scale $\sqrt{R E^{-1}}$. As $R$ and $n$ increase,
we may increase $y_l$ to obtain arbitrarily good exponential damping of the
AdS boundary effects from large scales.  Then, the wavefunctions
in the remaining regions are described by flat-space plane waves,
times the exponential envelope factor that essentially puts the plane waves
in finite volume.
As a result, all the important dynamics are taking place in a regime
where they can be described in terms of single-particle, flat-space
plane waves.

Now projecting (\ref{nextorderanzatz}) onto states with definite angular momentum, the correct flat-space states\footnote{
More precisely, Eq.~(\ref{eq:flatspacestates}) should be understood to
be true when it is acted on from the left by $\< 0 | \phi(x_1) \phi(x_2)$
for any $|x_1|,|x_2| \ll R$.} corresponding to two-particle primaries are
\be
|n,lJ\>_2 &=& \frac{|2p|^\frac{d-2}{2}}{(2\pi)^{d}\sqrt{2RE}}\int d\hat p\,Y_{lJ}(\hat p)\int d^d q\,f(q)|p+q\> | -p+q\>\qquad(n\gg 1,l),
\label{eq:flatspacestates}
\ee
where we have fixed the normalization by requiring that ${}_2\<n,lJ|n',l'J'\>_2=\de_{nn'}\de_{ll'}\de_{JJ'}$, which approaches $R^{-1}\de(E-E')\de_{ll'}\de_{JJ'}$ in the continuum limit.  Taking matrix elements of both sides, we find that
the leading large $n \gg l,1$ behavior of $\gamma(n,l)$ matches 
$\CM(s,t,u)^{d+1}_{\textrm{flat space}}$ after taking $2E = E_n k = (2\Delta + 2n)k,\, p^2 = n(n+2\Delta)k^2$
according to the relation
\be
\label{flatspacetheorem}
\g(n,l) &=& \frac{\vol(S^{d-2})}{(2\pi)^d}\frac{|p|^{d-2}}{8E}\int d\th \sin^{d-2}\th\,P_l^{(d)}(\cos\th) \CM(s,t,u)^{d+1}_\textrm{\textrm{flat space}}\nn\\
&=& \frac{\vol(S^{d-2})}{(4\pi)^d}\frac{(E_{n}^2-4\De^2)^{\frac{d-2}2}}{E_{n}}\int d\th \sin^{d-2}\th\,P_l^{(d)}(\cos\th)  \CM(s,t,u)^{d+1}_\textrm{\textrm{flat space}} ,
\ee
where we have introduced the angular polynomials $P_l^{(d)}(\cos\th)$, defined by
$P_l^{(d)}(\hat e\.\hat e') = \frac{1}{r_l}\vol(S^{d-1})\sum_J Y_{lJ}(\hat e)Y_{lJ}^*(\hat e')$,
where $r_l$ is the dimension of the spin-$l$ representation of $\SO(d)$. 
Finally we can invert this relation using the completeness relation
\be
\textrm{vol}(S^{d-2}) \sin^{d-2} \theta \sum_l r_l P_l(\cos \theta) P_l(\cos \theta') &=& \textrm{vol}(S^{d-1}) \delta(\theta-\theta')
\ee
to obtain the result given in Eq.~(\ref{eq:flatspacefromgamma}).  

\subsection{Examples}
\label{sec:flatoverlap}

\subsubsection{Example 1: $\phi^4$}

Now we will turn to a number of checks that the flat-space S-matrix
does indeed emerge from $\gamma(n,l)$ at large $n$, as described in Eqs.~(\ref{eq:flatspacefromgamma}) and (\ref{flatspacetheorem}). We will return
to units of $R=1$ for simplicity, since factors of $R$ cannot appear
in the flat space amplitude anyway.
Our first check is the simplest case, a $\mu^{3-d} \phi^4/4!$ interaction in
AdS$_{d+1}$, which has simply $\CM_\textrm{flat space}=\mu^{3-d}$.  We have essentially already computed the anomalous dimensions in Eq.~(\ref{eq:phi4anydim}); keeping track of the $O(1)$ coefficients,
one finds that the large $n,\Delta$ limit of $\gamma(n,0)$ is
\be
\gamma(n,0) &=& \mu^{3-d} \frac{\textrm{vol}(S^{d-1})}{8(2\pi)^d}
\left( \frac{ [n(n+2\Delta)]^{\frac{d-2}{2}}}{\Delta+n}\right).
\ee
We recognize the factor $(\Delta +n)$ as the energy $E$ in global
coordinates of each one-particle state, and similarly
the momentum is $p^2 = E^2 - m^2 = (\Delta + n)^2 - \Delta^2
 = n (n + 2 \Delta)$.  Finally, since $P_0^{(d)}(\cos\th)=1/r_0$, we see this exactly agrees with Eq.~(\ref{eq:flatspacefromgamma}).

\subsubsection{Example 2: $(\nabla \phi)^4$}

Our second example is $\mu^3 (\nabla \phi)^4/4!$ in AdS$_3$, where
explicit formulae for $\gamma(n,l)$ are known.  The flat-space
amplitude for this operator is
\be
\CM_{\textrm{flat space}} &=& \mu^3 \left( E^4 + \frac{2}{3} p^2 E^2 + p^4 \left(
\frac{1}{3} + \frac{2}{3} \cos^2 \varphi \right) \right).
\ee
The appropriate angular polynomials $P_l^{(2)}(\cos \varphi)$ in 2d
are $P_0^{(2)}(x)=1$ and $P_2^{(2)}(x) = 2 (x^2-\half )$
for $l=0$ and $l=2$, respectively. 
  Projecting $\CM_{\textrm{flat space}}(\cos\varphi)$ onto these
polynomials gives
\be
\CM_{\textrm{flat space}}(x) &=& \frac{\mu^3}{3}(3 E^4 + 2 E^2 p^2 + 2 p^4) P_0^{(2)}(x)
+ \frac{\mu^3}{3} p^4 P_2^{(2)}(x).
\label{eq:dphi4partialamp}
\ee
In order to bring the explicit expressions for $\gamma(n,0)$ and $\gamma(n,2)$ given 
in equations (\ref{eq:p6},\,\ref{eq:dphi4gamman2}) into this form, we can take the
leading terms at large
$\Delta, n$ and replace $\Delta \rightarrow \sqrt{E^2 - p^2},
n\rightarrow E - \sqrt{E^2 -p^2}$:
\be
6 \pi \mu^3
\gamma(n,0) &\stackrel{n, \Delta \gg 1}{\longrightarrow}&
\frac{7 n^4 + 28n^3 \Delta + 36 n^2 \Delta^2 + 16 n \Delta^3 + 3 \Delta^4}
{8(\Delta + n)} \rightarrow \frac{3 E^4 + 2 E^2 p^2 + 2 p^4}{8 E}, \\
6 \pi \mu^3
\gamma(n,2) &\stackrel{n,\Delta \gg 1}{\longrightarrow} &
\frac{n^2 (n+2\Delta)^2}{16(\Delta + n)} \rightarrow
\frac{ p^4}{16 E}.
\ee
This again agrees with the flat-space scattering partial wave amplitude
(\ref{eq:dphi4partialamp}) upon substituting into (\ref{eq:flatspacefromgamma}).

\subsubsection{Example 3: $\gamma(n,L)$ at maximum spin $L$}

Contact quartic interactions have a maximum spin $L$ for the
primary operators to which they contribute anomalous dimensions;
for instance, $(\nabla \phi)^4$ has $L=2$.  In \cite{polchinski},
a general form for such contributions $\tilde{\gamma}(n,L)$ for any operator
was obtained, and its dependence on $n$ and $\Delta$ is fixed by $L$.
Since the overall constant coefficient is undetermined and so cannot
be checked anyway, we will neglect
many proportionality constants in this subsection.
Consider first $d=2$,  where we may take the explicit expression
for $\tilde{\gamma}(n,L)$ in the large $n, \Delta$ limit, and replace them
by the appropriate energy and momentum as above:
\be
\tilde{\gamma}(n,L) &=& \pi \frac{\Gamma(n+L+1)\Gamma(2\Delta + n + L -1)
\Gamma(\Delta + n -\half) \Gamma(\Delta + n +L) }
{ 4 \Gamma(1+n) \Gamma(\Delta + n) \Gamma(\Delta + n +L +\half)
\Gamma(2\Delta + n -1)} \nn\\
&\stackrel{n,\Delta \gg 1}{\longrightarrow}& \frac{\pi}{4} \frac{[n(n+2\Delta)]^L}{\Delta+n}
\rightarrow \frac{\pi}{4} \frac{p^{2L}}{E} .
\ee
To compare this with flat space, we may use the amplitude for the
exchange of a heavy scalar with mass $M$
as a trick to generate the correct quartic-interaction
amplitude.  Specifically,  one may expand in $1/M$ and
take the leading non-zero term for a given $L$.
This then corresponds to the lowest-dimensional effective operator that
contributes to $\gamma(n,L)$.
 But, for that operator, $L$ is the largest spin that
gets a correction, so the leading non-zero term in the $1/M^2$ series
is $\tilde{\gamma}(n,L)$.

Let $\CA$ denote the amplitude for scalar exchange:
\be
\CA &\equiv& \mu^{5-d} \left( \frac{1}{s-M^2} + \frac{1}{t-M^2} + \frac{1}{u-M^2} \right),
\nn\\
s &=& (2E)^2, \ \ \ \ \ t,u = -2(E^2 \pm p^2 \cos \varphi -m^2).
\label{eq:scalarexchangeamplitude}
\ee
The angular polynomials in 2d are just $P_l(\varphi) = \cos(l\varphi)$, so we can project
the scalar exchange amplitude as
\be
2\int_0^\pi d\varphi  \cos (L \varphi) \CA
&\supset& - (1+(-1)^L) 2^{L+1}\mu^3 \int_0^\pi d \varphi \cos (L \varphi) \left(\frac{p^{2L} \cos^L\varphi}
{M^{2L+2}}\right)  \nn\\
&& \ = -(1+(-1)^L)2\pi \mu^3 \frac{p^{2L}}{M^{2L+2}} ,
\ee
which, after dividing by the normalization factor $\propto E$
from Eq.~(\ref{eq:flatspacefromgamma}),
matches the behavior from $\tilde{\gamma}(n,L)$ above.

Similarly,  in $d=4$, the large $\Delta, n$ limit of $\tilde{\gamma}(n,L)$ is
\be
\tilde{\gamma}(n,L) &\stackrel{n,\Delta \gg 1}{\propto}&
\frac{ [n(n+2\Delta)]^{L+1}}{ \Delta + n} \rightarrow
 \frac{p^{2(L+1)}}{E} .
\ee
The angular polynomials in 4d are $P_l(\varphi) \propto 
\frac{\sin((l+1)\varphi)}{\sin\varphi}$, and when we project the scalar
exchange amplitude onto them,
we find
\be
\int_0^\pi d \varphi \sin \varphi   \sin ((L+1) \varphi)  \CA
& \supset& - (1+(-1)^L)2^L \mu \int_0^\pi d \varphi \sin \varphi \sin ((L+1) \varphi) \left( \frac{p^{2L} \cos^L\varphi}
{M^{2L+2}} \right)  \nn\\
& = & - (1+(-1)^L)\frac{\pi \mu }{2} \frac{p^{2L}}{M^{2L+2}}.
\ee
In $d=4$, the wavefunction overlap factor is $\propto p^2 /E$,
which again accounts for the difference between $\gamma(n,L)$ and
the flat-space amplitude.

\subsubsection{Example 4: Scalar exchange in $d=2$}

Finally, we will compare the anomalous dimensions arising from 
the scalar exchange calculation done in section \ref{sec:schannel} with the flat-space amplitude.
In this section we will obtain from flat-space scattering the
complete scalar exchange contribution to $\gamma(n,0)$
at large $n$ and $\Delta$, but due to the difficulty of
evaluating the $t$- and $u$-channels in the CFT, we will be able
to check explicitly only the $s$-channel.  As a partial check
of the $t$- and $u$-channels, we will expand in inverse powers
of the exchanged scalar mass and compare to the known form
of $\gamma(n,0)$ from operators in the low-energy theory;
however, strictly speaking, this is a check only of the form
of $\gamma(n,0)$ at $n$'s below the dimension of the exchanged
scalar primary operator.  For simplicity we will focus on scalar exchange
in $d=2$.

Projecting the amplitude $\CA$ onto spin-$0$ modes in 2d,
we have
\be
\CA_0  &=& \mu^3 
\left( \frac{1}{4E^2 - M^2} \right)- 
\mu^3 \left(\frac{2}{M\sqrt{M^2+4p^2}} \right) .
\ee
We have explicitly separated out the first term in brackets as the
s-channel contribution to the spin-0 amplitude.
In order to compare to $\gamma(n,0)$, we need to take the large $n, \Delta$
limit from section \ref{sec:schannel}.  In $d=2$, the expression simplifies to
\be
\gamma(n,0)  &=&
   \frac{\mu^3 }{(4\pi)(\Delta_\chi + E_n -2)^2}
\left[ \frac{_3F_2\left( \left\{ 1, \frac{\Delta_\chi - E_n }{2} ,
\frac{ \Delta_\chi - E_n +2}{2} \right\}, \left\{ \frac{\Delta_\chi
+E_n}{2}, \frac{\Delta_\chi + E_n}{2} \right\}, 1 \right)}
{ E_n - \Delta_\chi} \right. \nn\\
&& \left.-
\frac{_3F_2\left( \left\{ 1, \frac{\Delta_\chi - E_n +2}{2} ,
\frac{ \Delta_\chi - E_n +2}{2} \right\}, \left\{ \frac{\Delta_\chi
+E_n}{2}, \frac{\Delta_\chi + E_n+2}{2} \right\}, 1 \right)}
{ E_n + \Delta_\chi} \right] .
\ee
To take the appropriate limit of the hypergeometric functions, we can
use the integral representation
\be
&&\phantom{1}_3F_2\left( \left\{ \delta, \delta+1, 1 \right\},
\left\{ M+2, M+2 \right\}, 1 \right) = \ \ \ \ \ \ \ \ \ \ \ \ \ \ \\
&& \ \ \ \ \ \
   \frac{\Gamma^2(M+2)}{\Gamma(M+2-\delta)\Gamma(M+1-\delta)
\Gamma(\delta+1)\Gamma(\delta)}  \int_0^1 dt \int_0^1 ds
\frac{t^\delta (1-t)^{M-\delta} s^{\delta-1} (1-s)^{M+1 - \delta} }
{1-st}, \nn 
\ee
taking $M+2=\frac{\Delta_\chi + E_n}{2}$ and $\delta = \frac{\Delta_\chi - E_n}{2}$.
The integral has a saddle point near $s,t = \delta/M$ at large
$\delta, M$.\footnote{The hypergeometric function ${}_3F_2$ is
analytic in all of its arguments, so we may perform the integral assuming
$\delta>0$, and then obtain the result at $\delta<0$ by analytic continuation.}  Around this point,  all the factors in the
integral except for $(1-st)$ simply contribute to cancel
the $\Gamma$-function prefactors, leaving behind just the value
$M^2/(M^2-\delta^2)$ of $(1-st)$. The same argument applies to both
${}_3F_2$'s.   Thus, we obtain
\be
\gamma(n,0) &\approx& \frac{ \mu^3}{(4\pi )( \Delta_\chi + E_n)^2}
\left[ \left( \frac{M^2}{M^2-\delta^2} \right) \frac{ 2\Delta_\chi}
{E_n^2 - \Delta_\chi^2} \right] =
\frac{\mu^3}{(8\pi) E_n (E_n^2 - \Delta_\chi^2)}.
\ee
This matches the s-channel amplitude using (\ref{flatspacetheorem}).

To consider the full amplitude with all channels included, we can
expand in $1/M$ and match to contributions from local operators in AdS$_3$.
We are not interested in operators of the form $(\phi (\d^2)^n \phi)
(\phi (\d^2)^m \phi)$, since these may be related to $\phi^4$ by
the equations of motion and therefore do not give new forms of
momentum-dependence.  To throw these out, we simply take
$s=2(E^2+p^2),t,u =-2(E^2 \pm p^2 \cos^2 \varphi)$ in Eq.
(\ref{eq:scalarexchangeamplitude}), i.e. we replace terms like 
$(p_i + p_j)^2$ with $2 p_i \cdot p_j$.  The remaining s-wave amplitude is then
\be
\CA_{0,\textrm{no }m^2} &=& \mu^3 \left( \frac{1}{2(E^2 + p^2) - M^2} \right) - \mu^3
\left( \frac{2}{\sqrt{(2E^2 + M^2)^2-4p^2 }} \right)
\ee
The first few  terms $\sim 1/M^2, 1/M^4, 1/M^6$ are just the
$\phi^4$, $(\nabla \phi)^2 \phi^2$, and $(\nabla \phi)^4$  contributions
that we have already checked.   The first new piece appears at
$O(1/M^8)$:
\be
\CA_{0,\textrm{no }m^2} &\supset& 8\frac{\mu^3}{M^8}
\left( E^6 - 3 E^4 p^2 - p^6 \right) .
\ee
This needs to match the contribution from the local operator
$(\nabla_\mu \nabla_\nu \phi)^2 (\nabla \phi)^2$.  Performing the explicit computation
(see appendix \ref{app:dphi4}), we obtain
\be
\gamma(n,0) &\propto&
\frac{\tilde{P}_8(n)}{(2\Delta + 2n -3)(2\Delta+2n-1)(2\Delta+2n+1)},
\ee
where $\tilde{P}_8(n)$ is an eighth-order polynomial in $n$, whose explicit form
is given in equation
(\ref{eq:p8}).
At large $n$, this expression for $\gamma(n,0)$ simply approaches
\be
\gamma(n,0) &\propto& \frac{E^6 - 3 E^4 p^2 -p^6}{E},
\ee
as we expect.

\sect{Conclusion}

In this paper we have argued that whenever the dilatation operator of a CFT admits a perturbative expansion, local interactions 
in global AdS provide a natural framework for organizing such a perturbation theory.  This is particularly true if for dimensions $\Delta < \Delta_{\rm Heavy}$ there are only a few single-trace primary operators, in which case AdS contains only a few fields.  This is analogous to the statement that whenever a Lorentz-invariant theory describes weakly-interacting particles, local interactions in Minkowski space are a convenient way of organizing perturbation theory.  This is especially useful if for energies
$E<M_{\rm Heavy}$ the Lorentz-invariant theory describes only a few particles.  However, there was an important difference in the Lorentz-invariant case.  Namely, for Lorentz-invariant theories there is a clear argument that perturbative unitarity 
(for all $E<M_{\rm Heavy}$) requires that the scale $\Lambda$ suppressing non-renormalizable interactions should satisfy $\Lambda \gtrsim M_{\rm Heavy}$.  
Such an argument was previously lacking in connecting theories in AdS to their dual CFTs.  Indeed, in AdS theories with suppressed higher-dimensional
bulk terms there appear to be non-trivial constraints on CFT correlation functions.   For example, in correlation functions involving conserved
currents, such as $T_{\mu\nu}$ or $J_{\mu}$, only certain polarization structures will appear -- those which follow from 
the lowest-dimension AdS bulk terms~\cite{Hofman:2008ar}.  From the CFT side it seemed strange that one polarization structure would be 
preferred over another.  It was therefore unclear whether CFTs needed to satisfy multiple independent requirements in order to have well-behaved 
AdS duals.

Our results suggest that there is a single requirement that naturally suppresses non-renormalizable interactions in the bulk.
Demanding perturbative unitarity for all operator dimensions $\Delta < \Delta_{\rm Heavy}$ places a bound on the scale
suppressing non-renormalizable AdS interactions of $\Lambda \gtrsim \Delta_{\rm Heavy}/R_{\rm AdS}$.  Moreover, the dimension of 
non-renormalizable operators is directly related to the rate of growth in the anomalous dimensions of double-trace operators $\gamma(n,l)$ as 
$n$ is increased.  It would be interesting to repeat our analysis for the case of a bulk gauge field or graviton, and verify that indeed requiring CFT perturbative 
unitarity up to some large dimension $\Delta_{\rm Heavy}$ leads to the suppression of certain polarization structures by appropriate powers of 
$\Delta_{\rm Heavy}$.  Extending the approach to fermions would also be desirable.

As supersymmetry did not appear to a play a role in the analysis, it may be possible that there are condensed matter
systems which enjoy conformal symmetry, for which the notion of an effective conformal theory might be useful.
In particular, if one could find a system with even a mild hierarchy in the dimension of operators, there might be 
a useful AdS dual which includes only the order parameter, a few relevant deformations, and possible conserved currents.
A possible way of detecting such a system could be to look for the suppression of particular polarization structures in correlation functions.
An outstanding question is the role of naturalness in determining which types of operators may have low dimensions 
in theories with a hierarchy.  A related question concerns the cosmological constant itself, and whether getting a large hierarchy 
is possible in non supersymmetric theories.  Finding a condensed matter system with a hierarchy might shed some light on these questions.

In addition to local bulk interactions, we have also considered probing bulk scalar exchange in AdS through the CFT anomalous dimensions 
$\gamma(n,l)$.  In doing so we have found evidence that these anomalous dimensions behave very much like S-matrix elements,  displaying a 
resonance-like behavior as $n$ passes through the dimension corresponding to the exchanged scalar.  More generally, for $n \gg 1$ we have shown that 
the anomalous dimensions simply turn into the partial wave expansion for the flat-space amplitudes of the higher-dimensional bulk theory.  It would be 
interesting to further explore this correspondence in other examples of CFTs where the anomalous dimensions are calculable.  It would also be very 
interesting to extend this analysis beyond tree level, where one could for example study the effect of renormalization group running in $n$.

It might also be useful to explore locality further by explicitly constructing bulk states localized in the extra dimension
$\rho$ and study their evolution.  By superimposing multiple-particle states (or considering operators without a definite
number of traces) one can also construct classical field states.  These might lead to a better understanding of
classical backgrounds such as small black holes at the center of AdS.  These and related investigations are left to future work.

\subsection*{Acknowledgments}

We would like to thank J. Penedones, J. Maldacena, and A. Cohen for
helpful conversations, and J. Polchinski, S. Giddings, and J. Kaplan 
for comments on the manuscript.  We also thank the Aspen Center for Physics
for its hospitality during the completion of this work.  ALF and EK are supported 
by DOE grant DE-FG02-01ER-40676 and NSF CAREER grant PHY-0645456, and EK is
supported also by an Alfred P. Sloan Fellowship.  DP and DSD are supported by 
the Harvard Center for the Fundamental Laws of Nature and by NSF grant PHY-0556111. 
\begin{appendix}

\sect{Check of $\gamma(n,0)$ for $(\nabla \phi)^4$ and $(\nabla_\mu\nabla_\nu \phi)^2 (\nabla \phi)^2$ }
\label{app:dphi4}

As a check of our methods, and to demonstrate them in a slightly more involved
example, we will show that they reproduce the scalar anomalous dimensions
$\gamma(n,0)$ calculated in \cite{polchinski} for a $(\nabla \phi)^4$ interaction in $d=2$.
From perturbation theory, we have
\be
\gamma(n,0) &=& \frac{1}{4! \mu^3} \int d^2x \sqrt{-g} {}_2\< n,0 |  (\nabla \phi)^4 | n,0 \>_2 \\
&=&  \frac{1}{6\mu^3} \int d^2x \sqrt{-g} \left[ \frac{1}{2} {}_2\< n,0 | (\nabla \phi)^2 | 0 \>
\< 0 | (\nabla \phi)^2 | n,0 \>_2  +
{}_2\< n,0 | \nabla^\mu \phi \nabla^\nu \phi | 0 \>
\< 0 | \nabla_\mu \phi \nabla_\nu \phi | n,0 \>_2 \right].\nn
\label{eq:dphi4split}
\ee
Using the identity
$\< 0 | (\nabla \phi)^2 | n,0 \>_2
= (\half m_n^2 - m^2) \< 0 | \phi^2 | n,0 \>_2$ with $m_n^2 = 4(\Delta+n) (\Delta + n -1)$,
the first term is easily reduced to an integral in terms of $\< 0 | \phi^2 |n,0 \>_2 = \frac{1}{\sqrt{2} \pi}(e^{it} \cos \rho)^{E_n}$,
which is straightforward to compute.  The second term is more complicated.
Since we are currently looking only at the dimensions of the scalar states $| n,0 \>_2$,
we want to decompose the operator $\nabla_\mu \phi \nabla_\nu \phi$ into
its scalar pieces.  The only primary wavefunctions
with two Lorentz indices that transform like scalars ($l=0$) come from
\be
\nabla_\mu \phi \nabla_\nu \phi &\supset&  \alpha g_{\mu\nu} \phi^2
+ \beta \nabla_\mu \nabla_\nu \phi^2.
\ee
  To determine the values of $\alpha$ and
$\beta$, we will manipulate $\nabla_\mu \phi \nabla_\nu \phi$ to
get linear combinations of just the scalar pieces.  That is,
$\nabla_\mu \phi \nabla_\nu \phi$ also contains a spin-2 piece
$H_{\mu\nu}$ $(H^\mu_\mu =0, \nabla_\mu H^\mu_{\ \nu} = 0)$
which needs to be projected out.\footnote{
Note that there is no possible spin-1 piece since the only vector
that could possibly enter is $\nabla_\mu \phi^2$, which has $l=0$ on primary
states.}
 The first projection is obtained
by taking the trace, which yields
\be
(\half m_n^2 - m^2)  &=& 3 \alpha + m_n^2 \beta .
\label{eq:proj1}
\ee
The second projection is obtained by acting
with $\nabla_\mu$, which picks out a different linear combination
of $\alpha$ and $\beta$,
\be
\frac{1}{4}m_n^2 \nabla_\nu \phi^2
 &=& (\alpha + (m_n^2 -2) \beta)  \nabla_\nu \phi^2 ,
\label{eq:proj2}
\ee
where we have used $[\nabla_\mu, \nabla_\nu] v^\mu =-2 v_\nu$.
Equations (\ref{eq:proj1}) and (\ref{eq:proj2}) have the solution
\be
\alpha &=& \frac{m_n^2 (m_n^2-4) - 4 m^2 (m_n^2-2) }
{ 8(m_n^2-3)},  \ \ \ \ \
\beta =  \frac{(4 m^2 + m_n^2)}{8(m_n^2-3)} .
\ee
The second term in our original integral then becomes
\be
|\<0| \nabla_\mu \phi \nabla_\nu \phi |n,0\>_2 |^2 &=& | g_{\mu \nu} \alpha \< 0 | \phi^2 | n,0 \>_2
 + \beta \nabla_\mu \nabla_\nu \< 0 | \phi^2 | n,0 \>_2 | ^2 \\
&=& \frac{1}{2\pi^2}
\left[(3 \alpha^2 + 2 \alpha \beta m_n^2)  \cos^{2 E_n} \rho
  +  \beta^2  \nabla^\mu \nabla^\nu (e^{-i t} \cos \rho)^{E_n}
 \nabla_\mu \nabla_\nu (e^{i t} \cos \rho)^{E_n} \right]  \nn.
\ee
Finally, integrating all terms over $\int_0^{2\pi} 
d\varphi \int_0^{\pi/2}  d\rho \sqrt{-g}$ we obtain
\be
6\mu^3 \pi \gamma(n,0) &=& \frac{\tilde{P}_6(n)}
{(2 n+2 \Delta -3)
(2 n+2 \Delta -1) (2 n+2
   \Delta +1)},
\ee
where $\tilde{P}_6(n)$ is the polynomial
\be
\tilde{P}_6(n) &=&
7 n^6+21 (2 \Delta -1) n^5+ \left(99 \Delta ^2-93 \Delta
   +16\right) n^4+(2 \Delta -1) \left(58 \Delta ^2-46 \Delta
   -3\right) n^3 \nn\\
&&+  \left(71 \Delta ^4-110 \Delta ^3+31 \Delta
   ^2+11 \Delta -5\right) n^2+\Delta ^3 \left(2 \Delta -1\right) \left(11
   \Delta -14\right) n \nn\\
&& +\frac{1}{4} \Delta ^3 (2 \Delta -3) \left(6
   \Delta ^2-5 \Delta +4\right) .
\label{eq:p6}
\ee
Compared to \cite{polchinski}, this agrees up to a term
proportional to the contribution from a $\int d^2x \sqrt{-g}\phi^4$ interaction in the bulk, 
which was intentionally dropped in their calculation.

For reference we will also write down the contribution to the 
spin-2 primary operators computed in \cite{polchinski}
\be
6\pi \mu^3 \gamma(n,2) &=& \frac{(n+1)(n+2)(n+\Delta)(n+\Delta+1)(n+2\Delta-1)(n+2\Delta)}
{2(2n+2\Delta-1)(2n+2\Delta+1)(2n+2\Delta+3)}.
\label{eq:dphi4gamman2}
\ee
Although we will not rederive this result here, it is straightforward to do so 
using the present method after determining the spin-2 primary wavefunctions in AdS${}_3$.

Finally, we have used similar manipulations to those above in order to compute
the scalar anomalous dimensions $\gamma(n,0)$ from the dimension-five
operator $(\nabla_\mu\nabla_\nu \phi)^2 (\nabla \phi)^2$ in $d=2$.  The
result we find is
\be
\gamma(n,0) &\propto& \frac{\tilde{P}_8(n)}
{(2\Delta + 2n -3)(2\Delta+2n-1)(2\Delta+2n+1)},
\ee
where $\tilde{P}_8(n)$ is an eighth-order polynomial in $n$, 
\be
\tilde{P}_8(n) &=&
-3 n^8+\left(-24 \Delta +12\right) n^7+ \left(-78 \Delta ^2+72
   \Delta -24\right) n^6 \nn\\
&& + \left(-132 \Delta ^3+162 \Delta ^2-108
   \Delta +30\right) n^5 \nn\\
&& + \left(-123 \Delta ^4+162 \Delta
   ^3-171 \Delta ^2+108 \Delta -15\right) n^4 \nn\\
&&
+\left(-60 \Delta
   ^5+54 \Delta ^4-108 \Delta ^3+144 \Delta ^2-36 \Delta -6\right)
   n^3 \nn\\
&&+\left(-11 \Delta ^6-24 \Delta ^5-9 \Delta ^4+88 \Delta
   ^3-33 \Delta ^2-12 \Delta +6\right) n^2 \nn\\
&& +\left(2 \Delta
   ^7-25 \Delta ^6+24 \Delta ^5+14 \Delta ^4-10 \Delta ^3\right)
   n \nn\\
&& +\frac{1}{4} \left(4\Delta ^8-28 \Delta ^7+45 \Delta ^6-14 \Delta
   ^5-22 \Delta ^4+24 \Delta ^3\right) .
\label{eq:p8}
\ee
Again this agrees with the results in~\cite{polchinski}, up to
subtracting off contributions from lower-dimensional operators.

\end{appendix}

\end{document}